\documentclass[reprint,aps,twocolumn]{revtex4-1}
\usepackage{amsmath}
\usepackage{graphicx}  
\usepackage{siunitx}   
\usepackage{url}  

\begin{document}

\title{Forces and Torques Near to Impact in the Golf Swing}
\author{Robert D. Grober}
\email[]{bob.grober@gmail.com}
\affiliation{Nissequogue, NY, 11780}

\date{\today}

\begin{abstract}
  Motivated by MacKenzie's observation of a negative force couple near
  to impact \cite{MacKenzie:2020a, MacKenzie:2016d}, this paper
  explores a model for how the golf club moves near to impact.  It
  assumes the golf club is moving as the distal arm of a double
  pendulum.  At impact the club head is moving straight down the
  target line, at its maximum speed, on a path with a specified radius
  of curvature.  From this model the forces and torques required to
  move the club near to impact are calculated.  The results are shown
  to be quantitatively consistent with data reported by MacKenzie to
  within a few percent.  The negative couple near to impact is a
  robust feature of this model, balancing the torque associated with
  the force that drives the center of mass of the golf club.  The
  negative couple allows the golfer to maintain a larger radius of
  curvature of the path of the club head as it moves through impact.
  Because the negative couple can also serve to reduce the rotational
  speed of the club, the presence of a negative couple at impact in
  the golf swing manifests a trade between distance and direction.
\end{abstract}


\maketitle
\tableofcontents
\newpage

\section{Acknowledgments}
This paper has benefited from the conversations, suggestions, and
thoughtful critique of many colleagues, including Grant Waite, Chris
Como, Sasho MacKenzie, Young-Hoo Kwon, Michael Finney, Bill Greenleaf,
Shawn Cox, Phil Cheetham, and Paul Wood.  The geometry defined in
Fig.~\ref{fig:dtl_family_of_positions}, which is the organizing
principle of the model explored in this paper, was inspired by ideas
expressed during a lecture by Michael Hebron.

\section{Summary for Golfers}
This section summarizes several of the salient points discussed in
this paper which may be of general interest to golfers. They are
presented in this summary without the mathematical detail provided in
the paper.

\begin{itemize}

\item This paper explores the forces and torques that move the club at
  impact. It assumes the club is moving as the distal arm of a double
  pendulum, as depicted in Fig.~\ref{fig:dtl_angle_defs_r2}.  At
  impact the club head is moving straight down the target line at
  maximum speed.

\item There is a geometry particular to the double pendulum which
  allows the club head to access points along the target line, as is
  shown in Fig. \ref{fig:dtl_family_of_positions}.  The length of the
  target line that is accessible depends on how far the golfer stands
  from the ball, but is typically 8-16 inches long, covering the
  distance from the middle of the stance to the forward foot.  In this
  geometry the hands are always ahead of the club head.  The path of
  the hands through this region is up and in.  This geometry is the
  organizing principle for the model explored in this paper.

\item In practice it is not possible to keep the club head moving on a
  straight line for an extended distance as it moves through impact at
  speed.  Rather, the club head moves on an arc, as depicted in
  Fig.~\ref{fig:dtl_calculated_path}.  It is possible to make the
  radius of curvature of the club head path sufficiently large that
  the deviation from a straight line is negligible for several inches
  before and after impact, Fig.~\ref{fig:dtl_calculated_path_zoom}.
  This allows some margin for error in the golf swing.

\item At impact the rotational speed of the proximal arm of the double
  pendulum (i.e. the shoulders, arms, and hands) is decreasing while
  the rotational speed of the distal arm (i.e. the club) is
  increasing, as can be inferred from Fig~\ref{fig:dtl_angles_dot}.
  This happens in a balanced way so as to allow the club head to move
  at maximum speed in a direction straight down the target line at
  impact.  The deceleration of the proximal arm in vicinity of impact
  is consistent with previous studies of the kinematic sequence
  \cite{Cheetham:2008a}.

\item As is shown in Fig.~\ref{fig:dtl_forces}, the force applied to
  the club by the golfer at impact is oriented in the general
  direction of the hub (i.e. the fixed pivot about which the proximal
  arm of the double pendulum rotates, which corresponds roughly to the
  middle of the sternum).  It is is dominated by the centripetal force
  needed to keep the center of mass of the club moving on an arc.
  Both the magnitude and orientation of the force are consistent with
  the inverse dynamics measurements of MacKenzie
  \cite{MacKenzie:2016d}.  The direction of the applied force at
  impact is an important result, and could be an organizing theme
  around which a golfer's biomechanics at impact are optimized.

\item This force applied by the golfer at impact results in a torque
  applied to the club which serves to increase the rotational speed of
  the club.  However, this torque also serves to decrease the radius
  of curvature of the path of the club head. To compensate for this,
  an additional torque is applied to the club so as to moderate the
  total torque without applying any additional net force.  The details
  of the balancing of these two torques are shown in
  Fig.~\ref{fig:dtl_torques_in_cm_frame}.  This additional torque
  takes the form of a force couple \cite{wiki:ForceCouple}, which can
  be though of as two forces, equal in magnitude, opposite direction,
  separated through a distance.  A force couple generates a torque,
  but does not accelerate the center of mass.
  
  This force couple has been measured by MacKenzie
  \cite{MacKenzie:2020a, MacKenzie:2016d} throughout the entire swing.
  It is negative within a few tens of milliseconds of impact, where it
  also acquires its largest magnitude.  This large, negative force
  couple in the vicinity of impact is ubiquitous among the golfers
  that have been measured.  It is surprising because a negative couple
  would reduce the rotational speed of the club, which seems contrary
  to the goals of most golfers.

  This paper shows that this negative couple in the vicinity of impact
  is a robust feature of the double pendulum model of the golf
  swing. It serves to reduce the total torque applied to the club,
  allowing the club head path to maintain a larger radius of curvature
  through the ball.  As such, the negative couple is a manifestation
  of the trade between distance and direction.

\item It remains the subject of future work to explain exactly how
  this negative force couple is generated.  Given that it occurs over
  an imperceptibly short period of time near to impact, and that
  nobody was aware of it before MacKenzie's experiments, this negative
  couple is possibly an involuntary feature of the body when the
  hands/wrists are rotating at very high speed.  If so, it suggests
  golfers have learned to incorporate this natural negative couple
  into their golf swings in a way which allows them to hit the ball
  straighter. Indeed, when training golfers it may be better to simply
  focus on the path of the club through the ball rather than trying to
  measure the force couple at impact.

\item Golfers are going to ask how this information can be used to
  improve their golf swing.  This question is best addressed by
  professional golf instructors.  However, it is interesting to point
  out that the deceleration of the hands and the orientation of the
  force at impact highlighted in this paper is reminiscent of an
  approach to training the golf swing named the `Rotor Method' that
  was pioneered by Nichols in the 1970s \cite{Fishman:1978b} and
  recently demonstrated in a video by Malaska \cite{Malaska:2018a}.
  Quoting from \cite{Fishman:1978b}, the downswing was characterized
  by `the explosive movement of the ... right side against the
  resistance of the left'.  This serves to enhance the deceleration of
  the torso/arms/hands at impact. At impact Nichols stressed ` ... the
  weight of the club head must go down the line until just after
  impact and then upward'.  Pulling the club upward just after impact
  serves to help the golfer orient the forces at impact towards the
  hub.  When done correctly, this style of `swing produces a very
  shallow arc resulting in long, thin divots'.  This is suggestive of
  the club head paths of Figs.~\ref{fig:dtl_calculated_path} and
  \ref{fig:dtl_calculated_path_zoom}.  Perhaps this training
  methodology from the 1970s can be adapted to the modern golf swing
  as a means of training the deceleration of the body and the
  hub-centric orientation of the applied force near to impact.
  
\end{itemize}

\section{Introduction}
This paper is motivated by the results of MacKenzie
\cite{MacKenzie:2020a,MacKenzie:2016d,MacKenzie:2016a,MacKenzie:2016b,MacKenzie:2016c},
Kwon \cite{Kwon:2017a} and Nesbit
\cite{Nesbit:2005b,Nesbit:2009a,Nesbit:2014a}, who have used 3-d
motion analysis of the golf club to infer the forces and torques
necessary to move the club throughout the swing.  A goal of this paper
is to understand the role of the negative couple in the immediate
vicinity of impact, as reported by MacKenzie \cite{MacKenzie:2016d}.

The golf swing has long been modeled as a double pendulum
\cite{Cochran:1968, Jorgensen:1970, Jorgensen:1994a}.  This paper
makes use of this model in the immediate vicinity of impact.  There
has been much discussion about the general applicability of the double
pendulum to the entire golf swing. For instance, it is known the hub
(i.e. the fixed pivot about which the proximal arm of the double
pendulum rotates) is not rigorously fixed throughout the entire swing
\cite{Jorgensen:1994a}, and there are claims the length of the
proximal arm can change significantly during the swing
\cite{Nesbit:2009a}.  This paper is focused on the dynamics in the
immediate vicinity of impact.  An explicit assumption of this paper is
that near to impact the hub is reasonably fixed and the proximal arm
is of constant length.  Under these conditions the double pendulum is
a good approximation.

The paper is divided into six sections.  The first section introduces
a geometry particular to the double pendulum in which the club can
access points along the target line.  The length of the target line
that is accessible depends on how far the golfer stands from the ball,
but generally extends from the middle of the stance out towards the
forward foot.  

The second section uses this geometry to constrain the dynamics of the
double pendulum so as to limit consideration to golf swings where the
club head reaches maximum speed as it moves down the target line at
impact on a path with a specified radius of curvature.

The third section begins with a derivation of the double pendulum
Lagrangian, done in the coordinate system used in this paper.  The
Lagrange equations of motion are used to calculate the external
torques required to drive the system at impact, given the constraints
in the second section.  It is in this section that the negative couple
reported by MacKenzie is found to be a robust feature of the model.

In the fourth section the equations of motion are used to simulate the
motion of the club in a region near to impact.  The external applied
torques are assumed constant throughout this region, equal to the
values required at impact.  Using inverse dynamics, similar to the
approach of MacKenzie \cite{MacKenzie:2016d}, Kwon \cite{Kwon:2017a}
and Nesbit \cite{Nesbit:2005b}, the simulated motion is used to
recover the forces and torques that move the club.

This fourth section provides the opportunity to look at the problem
from various different frames of reference, both inertial and
non-inertial.  This exercise serves to emphasize that the answer does
not depend on the frame of reference in which the problem is solved.
Hopefully, the discussions in this section can help to make clear some
of the issues associated with working in different frames of
reference \cite{Nesbit:2019_MR}.

The fifth section of the paper performs a search over the parameters
of the model to find the best fit to the forces and torques at impact
as reported by MacKenzie \cite{MacKenzie:2016d} for one particular
golfer.  It is demonstrated solutions to the model can be found which
agree quantitatively with MacKenzie's measurements to within a few
percent.

The final section of the paper speculates about various mechanisms
by which the negative force couple can be generated.

\section{Geometry of the Double Pendulum Near to Impact}

\begin{figure}
  \begin{center}
    \includegraphics[width=\columnwidth,angle=00]{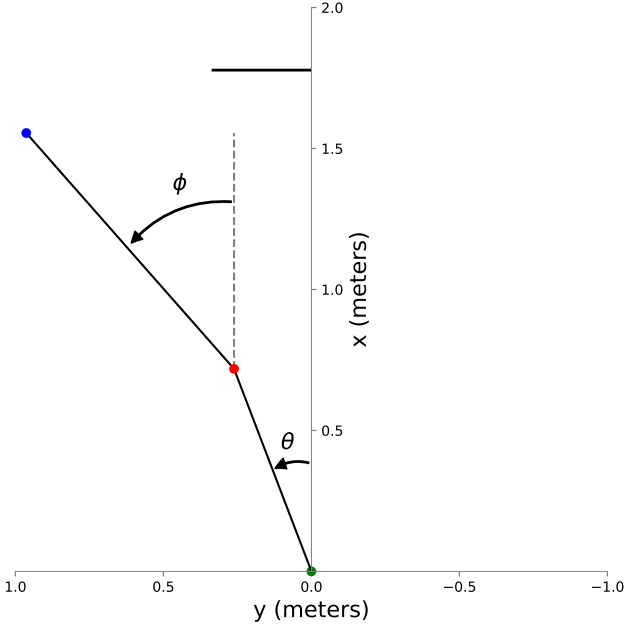}
    \caption{Geometry of the double pendulum, defining the angles
      $\theta$ and $\phi$.  The hub of the proximal arm is attached at
      the origin, indicated by the green circle, and is free to rotate
      about the origin.  The proximal arm is meant to approximate the
      shoulders/arms/hands of the golfer.  The distal arm is the golf
      club.  The hands attach to the golf club at the hinge indicated
      by the red circle.  The blue circle at the far end of the distal
      arm is the club head.}
    \label{fig:dtl_angle_defs_r1}
  \end{center}
\end{figure}

The coordinate system is shown in Fig.~\ref{fig:dtl_angle_defs_r1}.
The x-axis is perpendicular to the target line, while the y-axis is
oriented parallel to the target line.  The double pendulum consists of
two arms, a proximal arm of length $R_1$ and a distal arm of length
$R_2$.  The angles $\theta$ and $\phi$ describe the angle of the
proximal and distal arms relative to the x-axis.  The stationary end
of the proximal arm (i.e. the hub) is attached to the origin, but is
free to rotate about the origin.  The proximal arm is an approximation
to the shoulders/arms/hands.  The hinge between the proximal arm and
the distal arm is where the hands attach to the handle of the club.
The distal arm is the golf club.

The position $(x_1, y_1)$ of the far end of the proximal arm (i.e. the
hinge between the hands an the club handle) is

\begin{subequations}
\begin{align}
  x_1 = R_1 \cos \theta \\
  y_1 = R_1 \sin \theta
\end{align}
\end{subequations}
Similarly, the position $(x_2,y_2)$ of the end of the distal arm (i.e.
the club head) is

\begin{subequations}
\begin{align}
  x_2& = R_1 \cos \theta  + R_2 \cos \phi \\
  y_2& = R_1 \sin \theta + R_2 \sin \phi
\end{align}
\end{subequations}

\begin{figure}
  \begin{center}
    \includegraphics[width=\columnwidth,angle=00]{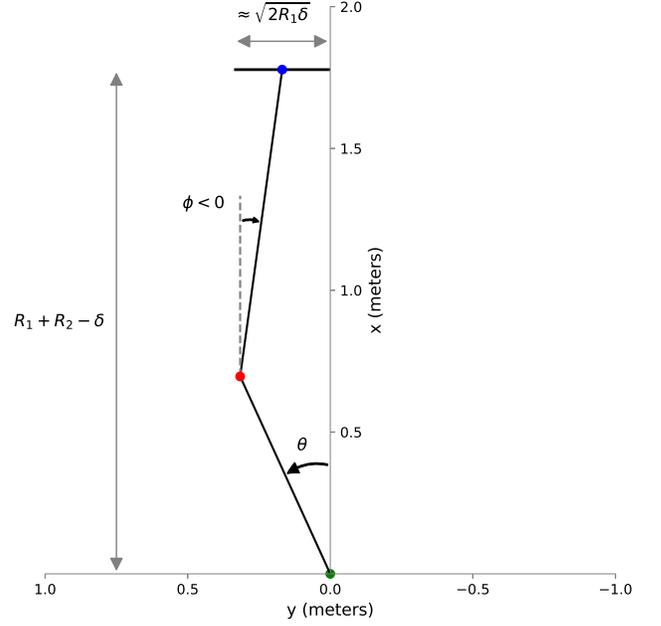}
    \caption{Geometry of the double pendulum, defining the angles
      $\theta$ and $\phi$, near to impact.  The club head, indicated
      by the blue circle, is fixed to the target line, which is a
      distance $R_1 + R_2 - \delta$ from the y-axis.  As is described
      in the text, the club head can access the target line from $y=0$
      thru $y \approx \sqrt{2 R_1 \delta}$.  In practice, this spans
      club positions from the middle of the stance out towards the
      forward foot.  Note that $\phi \le 0$ at all accessible points
      along the target line.}
    \label{fig:dtl_angle_defs_r2}
  \end{center}
\end{figure}

Assume the club is constrained to move straight down the target line,
parallel to the y-axis, a distance $x_0 = R_1 + R_2 - \delta$ from the
origin, where $\delta > 0$.  This is shown on the left side of
Fig.~\ref{fig:dtl_angle_defs_r2}. That $x_0 < R_1 + R_2$ allows the
club head to access a family of points straight down target line.  This
family is defined by the constraint

\begin{equation}
  R_1 + R_2 - \delta = R_1 \cos \theta + R_2 \cos \phi
\end{equation}

An additional constraint is that the hands should be slightly cocked
$(\theta - \phi)>0$.  Note that Jorgensen \cite{Jorgensen:1994a}
refers to the angle $\theta - \phi$ as $\beta$.  Subsequently, Nesbit
\cite{Nesbit:2005b} popularized the use of the Euler angle naming
convention $\alpha$, $\beta$, and $\gamma$ to describe rotation in the
swing plane, out of the swing plane, and about the axis of the shaft,
respectively.  This convention has become popular in golf teaching
circles, and thus the convention $\alpha = \theta - \phi$ is adopted
in this paper.

The final constraint is that the hands uncock as the club moves
towards impact

\begin{equation}
\frac{\delta \alpha}{\delta y} < 0
\end{equation}
These constraints yield a set of points along the target line,
starting at $(x,y) = (x_0,y_{min})$ through $(x,y) = (x_0,y_{max})$,
where $y_{min}=0$ and
$y_{max}=\sqrt{2 R_1 \delta - \delta^2} \approx \sqrt{2 R_1 \delta}$.
Note that at $y_{max}$, $\phi=0$.  For all other points along the
line, $\phi<0$.  Similarly, $\theta>0$ at all points along the target
line.

One can solve for $\theta$ and $\phi$ at all points where the club can
access the target line subject to these constraints, as follows.  The
parameters $x_0$ and $y_0$ describe the position of the club head on
the target line,

\begin{subequations}
\begin{align}
  x_0& = R_1 \cos \theta  + R_2 \cos \phi \\
  y_0& = R_1 \sin \theta + R_2 \sin \phi
\end{align}
\end{subequations}
Eliminate $\phi$ from these coupled equations by squaring and adding
together, 

\begin{subequations}
\begin{align}
  \bigl(R_2 \cos \phi \bigr)^2 & = \bigl(x_0 - R_1 \cos \theta \bigr)^2  \\
  \bigl(R_2 \sin \phi \bigr)^2 & = \bigl(y_0 - R_1 \sin \theta \bigr)^2
\end{align}
\end{subequations}
yielding

\begin{equation}
2 x_0 R_1 \cos \theta + 2 y_0 R_1 \sin \theta = x_0^2 + y_0^2 + R_1^2 - R_2^2 
\end{equation}
Simplify by defining the parameters $A = x_0^2 + y_0^2 + R_1^2 -
R_2^2$, $B = 2 x_0 R_1$, and $C = 2 y_0 R_1$.  Reduce to terms only
involving $\sin \theta$ by again taking the square

\begin{equation}
\bigl( B \cos \theta \bigr)^2 = B^2 \bigl( 1 - \sin^2 \theta \bigr) =  \bigl( A - C \sin \theta \bigr)^2 
\end{equation}
which yields a quadratic equation in $\sin \theta$

\begin{equation}
\bigl( B^2 + C^2 \bigr) \sin^2 \theta  - 2 A C \sin \theta +  \bigl(
A^2 - B^2 \bigr) = 0
\end{equation}
Changing parameters again, this time to $a = B^2 + C^2$, $b = - 2 A C$ and $c = A^2
- B^2$, and solving for $\sin \theta$,

\begin{equation}
\sin \theta  = \frac{-b + \sqrt{b^2 - 4 a c}}{2 a}
\end{equation}
Then use $y_0 = R_1 \sin \theta + R_2 \sin \phi$ to solve for
$\sin \phi$.  

\begin{figure}[!htb]
  \begin{center}
    \includegraphics[width=\columnwidth,angle=00]{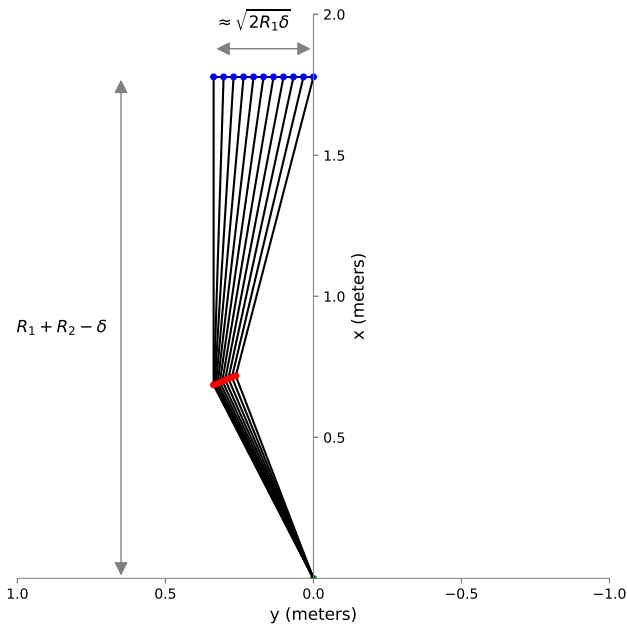}
    \caption{The family of orientations of the double pendulum for
      which the club head, shown as blue dots, can access points along
      the target line.  The hands, shown as red dots, are always ahead
      of the club head.  The hands have gone past the bottom of their
      arc, and as a result are traveling up and in relative to the
      target line}
    \label{fig:dtl_family_of_positions}
  \end{center}
\end{figure}

The resulting family of orientations of the double pendulum for which
the club head can access the target line is shown in
Fig.~\ref{fig:dtl_family_of_positions}.  The length of the distal arm
(i.e. the club) $R_2 = \SI{1.092}{\meter}$, consistent with the value
used in \cite{Nesbit:2009a}.  The length of the proximal arm
$R_1 = 0.7 R_2$ for no particular reason other than the aspect ratio
looks about correct.
$R_1 + R_2 = \SI{1.856}{\meter} \approx \SI{73}{in}$.  Finally,
$\delta$ is chosen to be $\SI{7.84}{\cm} \approx \SI{3}{in}$, which
makes $R_1 + R_2 - \delta = \SI{1.778}{\m} = \SI{70.0}{in}$.  The
length $\sqrt{2 R_1 \delta} \approx \SI{13.3}{in}$, which spans the
distance from the center of the stance out towards the the forward
foot (i.e. left foot for a right handed golfer).  The complete set of
model parameters used throughout this paper are provided in
Appendix~\ref{appdx:ap1}.

An important feature of Fig.~\ref{fig:dtl_family_of_positions} is the
motion of the hands near to impact.  The hands are always ahead of the
club head.  The hands have gone past the bottom of their arc, and as a
result are traveling up and in relative to the target line.

\section{Dynamics of the Double Pendulum Near to Impact}
The next step is to use this geometry to put constraints on the first and
second time derivatives of $\theta$ and $\phi$.

Start by considering the velocity of the club head as it moves down the
line.  It is useful to organize the expressions for position as a
matrix equation,
\begin{gather}
  \begin{bmatrix} x \\ y \end{bmatrix} =
  \begin{bmatrix}
    \cos \theta & \cos \phi \\
    \sin \theta & \sin \phi
  \end{bmatrix}
    \begin{bmatrix} R_1 \\ R_2 \end{bmatrix}
\end{gather}
Taking derivatives of the equations above,

\begin{gather}
  \begin{bmatrix} \dot{x} \\ \dot{y} \end{bmatrix} =
  \begin{bmatrix}
    -\sin \theta & -\sin \phi \\
    \cos \theta & \cos \phi
  \end{bmatrix}
    \begin{bmatrix}
    R_1 \dot{\theta} \\
    R_2 \dot{\phi}
  \end{bmatrix}
\end{gather}
The club is constrained to move straight down the line at impact,
so $\dot{x}_0 = 0$ and $\dot{y}_0 = v$.

\begin{gather}
  \begin{bmatrix} 0 \\ v \end{bmatrix} =
  \begin{bmatrix}
    -\sin \theta & -\sin \phi \\
    \cos \theta & \cos \phi
  \end{bmatrix}
    \begin{bmatrix}
    R_1 \dot{\theta} \\
    R_2 \dot{\phi}
  \end{bmatrix}
\end{gather}
Solving for $\dot{\theta}$ and $\dot{\phi}$ requires inverting the
matrix

\begin{gather}
  \begin{bmatrix}
    -\sin \theta & -\sin \phi \\
    \cos \theta & \cos \phi
  \end{bmatrix} ^{-1} =
  \frac{1}{\sin \alpha}
  \begin{bmatrix}
    -\cos \phi & -\sin \phi \\
    \cos \theta & \sin \theta
  \end{bmatrix}
\end{gather}
where $\alpha = \theta - \phi$ and
$\sin\alpha = \sin\theta \cos\phi - \sin\phi \cos\theta$.  Thus

\begin{multline}
   \begin{bmatrix}
    R_1 \dot{\theta} \\
    R_2 \dot{\phi}
  \end{bmatrix} =
  \frac{1}{\sin \alpha}
  \begin{bmatrix}
    -\cos \phi & -\sin \phi \\
    \cos \theta & \sin \theta
  \end{bmatrix}
  \begin{bmatrix} 0 \\ v \end{bmatrix} \\ =
  \frac{v}{\sin \alpha}
  \begin{bmatrix}
    - \sin \phi \\ \sin \theta
  \end{bmatrix}
\end{multline}
Through this entire region $\phi < 0$, and $\alpha > 0$.  Therefore,
both $\dot{\theta} > 0$ and $\dot{\phi} > 0$.

\begin{figure}[!htb]
  \begin{center}
    \includegraphics[width=\columnwidth,angle=00]{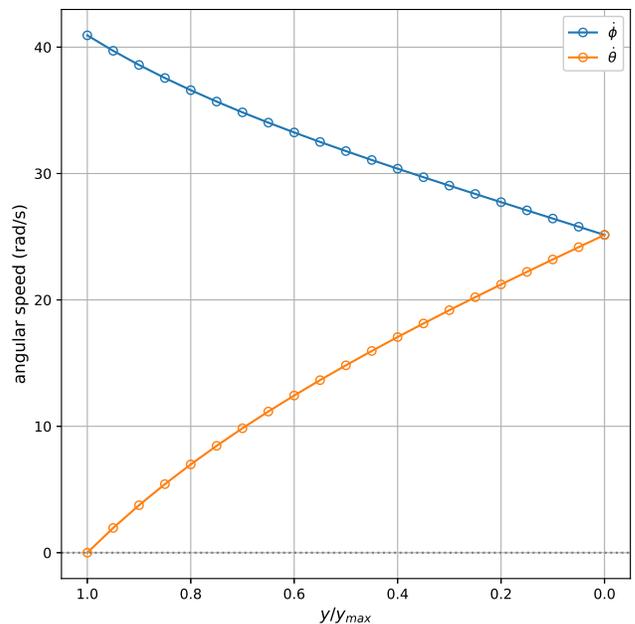}
    \caption{Angular speeds $\dot{\phi}$ and $\dot{\theta}$ for points
      along the target line calculated for a club head speed of
      $\SI{44.7}{\meter/\second}$ (i.e. 100 mph).  To keep the club on
      the target line, the club rotation (i.e. distal arm of the
      double pendulum) accelerates and the shoulder/arm/hand rotation
      (i.e. proximal arm of the double pendulum) decelerates.  The
      complete set of model parameters used throughout this paper are
      provided in Appendix~\ref{appdx:ap1}.}
    \label{fig:dtl_angles_dot}
  \end{center}
\end{figure}

Fig~\ref{fig:dtl_angles_dot} shows $\dot{\theta}$ and $\dot{\phi}$ for
points along the target line from $y=0$ thru $y=y_{max}$.  The
abscissa is setup to align with the image in
Fig.~\ref{fig:dtl_family_of_positions}. These angular speeds are
calculated for the case of a club speed of $\SI{44.7}{\meter/\second}$
(i.e. 100 mph).  Note that at $y=0$, $\dot{\theta} = \dot{\phi}$, and
thus the proximal and distal arms move together.  Out near $y=y_{max}$
at the end of the accessible target line, $\dot{\theta}=0$ and all
motion of the club head is related to $\dot{\phi}$.

The acceleration is constrained such that the club comes to its
maximum speed at impact, $\ddot{y}_0=0$.  Additionally, the club
travels from inside the line to inside the line, so at impact
$\ddot{x}_0<0$.  In principle, the magnitude $|\ddot{x}_0|$ should be
as small as possible so that the club head travels a reasonably straight
path down the target line.  In practice, it requires larger forces and
torques as the golfer makes $|\ddot{x}_0|$ smaller, and it becomes
impractical to get the club head to travel perfectly straight down the
target line for an extended distance at speed.  But, as will be shown
below, the resulting radius of curvature of the club path can be
sufficiently large that the club path is reasonably approximated as a
straight line near to impact.

The radius of curvature of the club head path at impact is given by
the expression $R_c = \dot{y}_0^2/\ddot{x}_0$ \cite{Anton:2009a}.  As
will be shown, it is useful to parameterize $R_c$ in terms of the
distance of the hub from the target line, $R_1 + R_2 - \delta$.  In
particular, define $R_c$ in terms of the parameter $\xi$ such that
$\dot{y}_0^2/|\ddot{x}_0| = \xi (R_1 + R_2 - \delta)$.  The condition
$\xi=1$ corresponds to the case when the path is approximated by the
perimeter of a circle of radius $(R_1 + R_2 - \delta)$.  Expressing
$\ddot{x}_0$ in terms of $\xi$,
\begin{equation}
\ddot{x}_0 = - \frac{\dot{y}_0^2}{\xi (R_1 + R_2 - \delta)}.
\end{equation}

The second derivatives $\ddot{x}$ and $\ddot{y}$ are given by the expression
\begin{multline}
  \begin{bmatrix} \ddot{x} \\ \ddot{y} \end{bmatrix} =
  \begin{bmatrix}
    -\sin \theta & -\sin \phi \\
    \cos \theta & \cos \phi
  \end{bmatrix}
  \begin{bmatrix}
    R_1 \ddot{\theta} \\
    R_2 \ddot{\phi}
  \end{bmatrix} \\
  -
  \begin{bmatrix}
    \cos \theta & \cos \phi \\
    \sin \theta & \sin \phi
  \end{bmatrix}
  \begin{bmatrix}
    R_1 \dot{\theta}^2 \\
    R_2 \dot{\phi}^2
  \end{bmatrix}
\end{multline}
where we impose the condition.

\begin{gather}
  \begin{bmatrix} \ddot{x}_0 \\ \ddot{y}_0 \end{bmatrix} =
  \begin{bmatrix} - \frac{\dot{y}^2}{\xi (R_1 + R_2 - \delta)} \\ 0 \end{bmatrix}
\end{gather}
Solving for $R_1 \ddot{\theta}$ and $R_2 \ddot{\phi}$ using the same
matrix inversion from above


\begin{multline}
   \begin{bmatrix}
    R_1 \ddot{\theta} \\
    R_2 \ddot{\phi}
  \end{bmatrix} =
  \frac{1}{\sin \alpha}
  \begin{bmatrix}
    -\cos \phi & -\sin \phi \\
    \cos \theta & \sin \theta
  \end{bmatrix}
  \begin{bmatrix} \ddot{x}_0 \\ \ddot{y}_0 \end{bmatrix} \\
  + \frac{1}{\sin \alpha}
  \begin{bmatrix}
    -\cos \alpha & -1 \\
    1 & \cos \alpha
  \end{bmatrix}
  \begin{bmatrix}
    R_1 \dot{\theta}^2 \\
    R_2 \dot{\phi}^2
  \end{bmatrix}
\end{multline}
With these equations, $(\theta$, $\dot{\theta}$, $\ddot{\theta})$, and
$(\phi$, $\dot{\phi}$, $\ddot{\phi})$ are fully specified at impact.

\begin{figure}[!htb]
  \begin{center}
    \includegraphics[width=\columnwidth,angle=00]{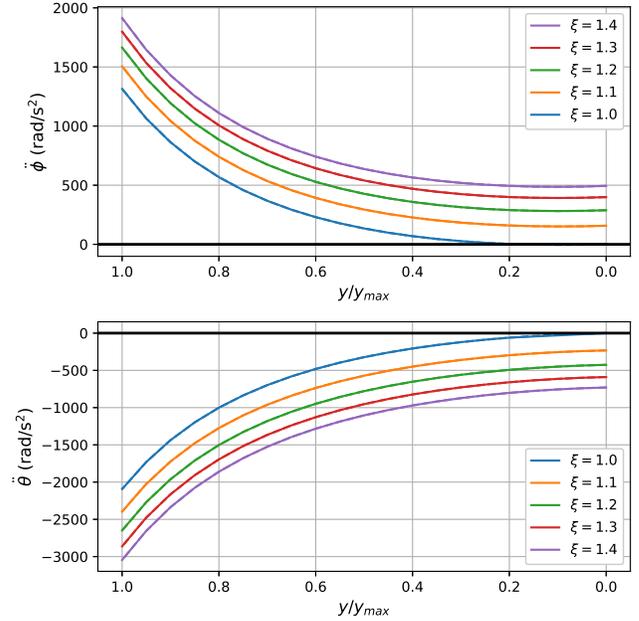}
    \caption{Angular acceleration $\ddot{\phi}$ and $\ddot{\theta}$
      for points along the target line for values $\xi \ge 1$.  There
      is a general trend that larger radius of curvature requires
      acceleration of larger magnitude.}
    \label{fig:dtl_angles_ddot_xi_gt1}
  \end{center}
\end{figure}

\begin{figure}[!htb]
  \begin{center}
    \includegraphics[width=\columnwidth,angle=00]{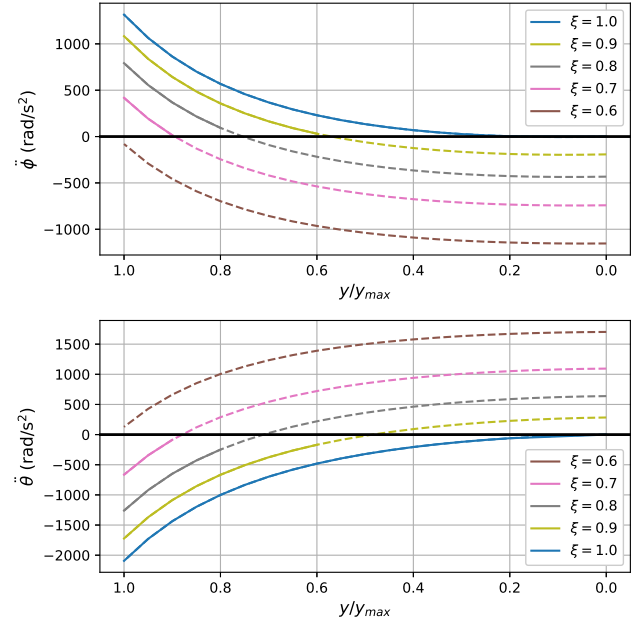}
    \caption{Angular acceleration $\ddot{\phi}$ and $\ddot{\theta}$
      for points along the target line for values $\xi \le 1$.  The
      dashed lines occur when either $\ddot{\phi}<0$
      (i.e. deceleration of rotation of the club) or $\ddot{\theta}>0$
      (i.e. acceleration of the rotation of the arms/hands).  As such,
      the dashed lines correspond to solutions which are not likely to
      be realized.}
    \label{fig:dtl_angles_ddot_xi_lt1}
  \end{center}
\end{figure}

Shown in Fig.~\ref{fig:dtl_angles_ddot_xi_gt1} are $\ddot{\phi}$ and
$\ddot{\theta}$ as a function of the distance along the target line
for various values of $\xi \ge 1$.  Again, the abscissa is setup to
align with the image in Fig.~\ref{fig:dtl_family_of_positions}.  Note
that at $y/y_{max} = 0$ and for $\xi = 1.0$,
$\ddot{\theta} = \ddot{\phi} \approx 0$.  For larger values of
$y/y_{max}$ the magnitude of the required angular acceleration
increases, with $\phi$ accelerating and $\theta$ decelerating.  The
deceleration of $\theta$ and the acceleration of $\phi$ near to impact
is consistent with previous studies of the kinematic sequence
\cite{Cheetham:2008a}.

Fig.~\ref{fig:dtl_angles_ddot_xi_lt1} details $\ddot{\phi}$ and
$\ddot{\theta}$ as a function of the distance along the target line
for various values of $\xi \le 1$.  Here the curves have a dashed
region and a solid region.  The dashed regions occur when either
$\ddot{\phi}<0$ (i.e. deceleration of rotation of the club) or
$\ddot{\theta}>0$ (i.e. acceleration of the rotation of the
arms/hands).  Neither solution is likely to be realized in practice.
The solid regions correspond to $\ddot{\phi} \ge 0$ and
$\ddot{\theta} \le 0$, and are the solutions which have a better
chance of matching what is realized in actual golf swings.  Note that
for the case of the solid lines, the magnitude of the acceleration
gets smaller at smaller values of $\xi$, which corresponds to the club
path through impact being more curved.

\section{\label{lagrange}The Lagrangian of the Double Pendulum}
The the double pendulum was originally used as a model for the golf
swing by Cochran and Stobbs \cite{Cochran:1968}.  The Lagrangian of
the double pendulum and its application to the dynamics of the golf
swing was subsequently pioneered by Jorgensen \cite{Jorgensen:1970}.
In this section, the Lagrangian is re-derived using the coordinate
system of this paper.

\subsection{Kinetic Energy}
The Lagrangian of a rigid body can be calculated as the difference
between the kinetic energy and potential energies \cite{Goldstein:1981}.
Thus, the first step is to define the kinetic energy of the moving
parts in the double pendulum.

Start by considering the proximal arm. Let $r_1$ denote the distance
along the arm.  The velocity of a point along the proximal arm is

\begin{subequations}
\begin{align}
  \dot{x_1} &= - r_1 \dot{\theta} \sin{\theta} \\
  \dot{y_1} &= r_1 \dot{\theta} \cos{\theta}
\end{align}
\end{subequations}
The square of the velocity is

\begin{equation}
v_1^2 = \dot{x_1}^2 + \dot{y_1}^2 = r_1^2 \dot{\theta}^2
\end{equation}

The kinetic energy is calculated by integrating the
local kinetic energy over the entire proximal arm.

\begin{equation}
KE_1 = \frac{1}{2} \int{dm_1 v_1^2}
\end{equation}
Defining the linear mass density $\rho_1(r_1)$ such that
$dm = dr_1 \rho_1(r_1)$, the integral becomes

\begin{equation}
KE_1 = \frac{1}{2} \int_0^{R_1}{dr_1 \rho_1 r_1^2 \dot{\theta}^2 }
\end{equation}
The integral of the linear mass density is just the mass,

\begin{equation}
M_1 = \int_0^{R_1}{dr_1 \rho_1}
\end{equation}
As such, $\rho_1/M_1$ is a probability density,
\begin{equation}
1 = \int_0^{R_1}{dr_1 \frac{\rho_1}{M_1}}
\end{equation}
With this interpretation, the integral over $r_1^2$ is the
second moment,
\begin{equation}
<R_1^2> = \int_0^{R_1}{dr_1 \frac{\rho_1}{M_1} r_1^2}
\end{equation}
The kinetic energy can then be parameterized as

\begin{equation}
KE_1 = \frac{1}{2} M_1 <R_1^2> \dot{\theta}^2
\end{equation}

Now consider the distal arm.  Define $r_2$ to be the distance along
the distal arm.  The velocity of a point along the distal arm is

\begin{subequations}
\begin{align}
  \dot{x_2} &= - R_1 \dot{\theta} \sin{\theta} - r_2 \dot{\phi} \sin{\phi} \\
  \dot{y_2} &= \; R_1 \dot{\theta} \cos{\theta} + r_2 \dot{\phi} \cos{\phi}
\end{align}
\end{subequations}
The square of the velocity is

\begin{equation}
  v_2^2 = \dot{x_2}^2 + \dot{y_2}^2
\end{equation}

\begin{equation}
  v_2^2 = R_1^2 \dot{\theta}^2 + r_2^2 \dot{\phi}^2 + 2 R_1 r_2 \dot{\theta} \dot{\phi} \bigl( \cos{\theta} \cos{\phi} + \sin{\theta} \sin{\phi} \bigr)
\end{equation}
which simplifies to 

\begin{equation}
  v_2^2 = R_1^2 \dot{\theta}^2 + r_2^2 \dot{\phi}^2 + 2 R_1 r_2 \dot{\theta} \dot{\phi} \cos{(\theta - \phi)}
\end{equation}

Once again, define the kinetic energy of the distal arm as an integral
of the local kinetic energy over the entire distal arm

\begin{equation}
KE_2 = \frac{1}{2} \int{dm_2 v_2^2}
\end{equation}

Defining the linear mass density of the distal arm, $\rho_2$, and
using the definitions of the first and second moments of the distal arm,

\begin{equation}
<R_2> = \int_0^{R_2}{dr_2 \frac{\rho_2}{M_2} r_2}
\end{equation}

\begin{equation}
 <R_2^2> = \int_0^{R_2}{dr_2 \frac{\rho_2}{M_2} r_2^2}
\end{equation}
the following expression for the kinetic energy of the distal arm is
obtained,

\begin{multline}
  KE_2 = \frac{1}{2} M_2 \bigl(R_1^2 \dot{\theta}^2 + <R_2^2>
  \dot{\phi}^2 \\ + 2 R_1 <R_2> \dot{\theta} \dot{\phi} \cos{(\theta - \phi)} \bigr)
\end{multline}

The kinetic energy of the entire double pendulum is then $KE = KE_1 +
KE_2$,

\begin{multline}
  KE = \frac{1}{2} \bigl(M_1 <R_1^2> + M_2 R_1^2 \bigr) \dot{\theta}^2
  + \frac{1}{2} M_2 <R_2^2> \dot{\phi}^2 \\ + M_2 R_1 <R_2> \dot{\theta} \dot{\phi} \cos{(\theta - \phi)}
\end{multline}

It is useful to define the following parameters
\begin{subequations}
\begin{align}
A & = M_1<R_1^2> + M_2 R_1^2 \\
B & = M_2<R_2^2> \\
C & = M_2 R_1 <R_2> \\
\alpha & = \theta - \phi  
\end{align}
\end{subequations}
With these definitions, the kinetic energy simplifies to

\begin{equation}
  KE = \frac{1}{2} A \dot{\theta}^2 + \frac{1}{2} B \dot{\phi}^2 + C \dot{\theta} \dot{\phi} \cos{\alpha}
\end{equation}

The values of the parameters in $A$, $B$, and $C$ which are
used in subsequent calculations in this paper are listed in
Appendix~\ref{appdx:ap1}.

\subsection{Potential Energy}
We will want to apply some external torques to the system.  These
torques are better described as force couples \cite{Simon:1971}, where
a force couple $K$ can be thought of as the torque obtained by two
forces, equal in magnitude $F$ but opposite in direction, acting at
two different points separated by a distance $d$. The net force is
zero, so the couple does not accelerate the center of mass.  The net
torque is $K = F d$, and results in rotation about the center of mass.

Assume a couple of constant magnitude $K_\theta$ is applied at the hub
and has the orientation such that it increases the angle $\theta$.
The potential energy for this couple is

\begin{equation}
PE_1 = - K_\theta \theta
\end{equation}
Similarly, another couple, $K_\alpha$ is applied at the hinge between the
two arms.  It is applied such that it increases the angle $\phi$
relative to $\theta$, and thus decreases $\theta - \phi$.  The
potential energy associated with a constant version of this couple is then

\begin{equation}
PE_2 = K_\alpha (\theta - \phi)
\end{equation}
As will be shown below, $K_\alpha$ corresponds to the couple reported
by MacKenzie \cite{MacKenzie:2016d}.

The total potential energy is
\begin{equation}
PE = - K_\theta \theta + K_\alpha (\theta - \phi)
\end{equation}
The resulting Lagrangian $L$ is

\begin{equation}
L = \frac{1}{2} A \dot{\theta}^2 + \frac{1}{2} B \dot{\phi}^2 + C
\dot{\theta} \dot{\phi} \cos{\alpha} + K_\theta \theta - K_\alpha (\theta - \phi)
\end{equation}

\subsection{Equations of Motion}
The Lagrangian is of the form $L(x_i, \dot{x_i})$, where $i$ ranges
over the independent coordinates, in this case $\theta$ and $\phi$.
The associated equation of motion for each coordinate is given by
\cite{Goldstein:1981},

\begin{equation}
\frac{d}{dt} \frac{\delta L}{\delta \dot{x_i}} - \frac{\delta
  L}{\delta x_i} = 0
\end{equation}
The equation of motion associated with $\theta$ is

\begin{equation} \label{eom_theta}
  A \ddot{\theta} + C \ddot{\phi}
  \cos{(\theta - \phi)} + C \dot{\phi}^2 \sin{(\theta -
    \phi)} = K_\theta - K_\alpha
\end{equation}
Similarly, the equation of motion associated with $\phi$ is
\begin{equation} \label{eom_phi}
B \ddot{\phi} + C \ddot{\theta}
\cos{(\theta - \phi)} - C \dot{\theta}^2 \sin{(\theta -
  \phi)} = K_\alpha
\end{equation}
These are the two equations of motion which govern the motion of the
double pendulum subject to couples $K_\theta$ and $K_\alpha$.  Given
initial conditions ($\theta_0$, $\dot{\theta}_0$) and ($\phi_0$,
$\dot{\phi}_0$), and the couples $K_\theta$ and $K_\alpha$, the
equations of motion can be solved for $\theta(t)$ and $\phi(t)$.

\subsection{Solving for Couples}
Consider the situation at impact.  The values ($\theta_0$,
$\dot{\theta}_0$, $\ddot{\theta}_0$) and ($\phi_0$, $\dot{\phi}_0$,
$\ddot{\phi}_0$) are known from the constraint that at impact the club
moves down the target line at peak speed on a path with a specified
radius of curvature.  In this section the equations of motion are
inverted to solve for the values of $K_\theta$ and $K_\alpha$ that are
consistent with this condition.

Start with the equations of motion above, now written in matrix
notation

\begin{multline}
  \begin{bmatrix}
    A & C \cos{\alpha} \\
    C \cos{\alpha} & B
  \end{bmatrix}
  \begin{bmatrix}
    \ddot{\theta} \\
    \ddot{\phi}
  \end{bmatrix} +
  \begin{bmatrix}
    0 & C \sin{\alpha} \\
    -C \sin{\alpha} & 0
  \end{bmatrix}
  \begin{bmatrix}
    \dot{\theta}^2 \\
    \dot{\phi}^2
  \end{bmatrix} \\ = 
  \begin{bmatrix}
    1 & -1 \\
    0 & 1
  \end{bmatrix}
  \begin{bmatrix}
    K_\theta \\
    K_\alpha
  \end{bmatrix}
\end{multline}
Invert the matrix in front of $K_\theta$ and $K_\alpha$,
\begin{gather}
  \begin{bmatrix}
    1 & -1 \\
    0 & 1
  \end{bmatrix}^{-1} = 
  \begin{bmatrix}
    1 & 1 \\
    0 & 1
  \end{bmatrix}
\end{gather}
Solve for $K_\theta$ and $K_\alpha$,
\begin{multline}
  \begin{bmatrix}
    K_\theta \\
    K_\alpha
  \end{bmatrix} =
  \begin{bmatrix}
    1 & 1 \\
    0 & 1
  \end{bmatrix}
  \begin{bmatrix}
    A & C \cos{\alpha} \\
    C \cos{\alpha} & B
  \end{bmatrix}
  \begin{bmatrix}
    \ddot{\theta} \\
    \ddot{\phi}
  \end{bmatrix} \\ +
  \begin{bmatrix}
    1 & 1 \\
    0 & 1
  \end{bmatrix}
  \begin{bmatrix}
    0 & C \sin{\alpha} \\
    -C \sin{\alpha} & 0
  \end{bmatrix}
  \begin{bmatrix}
    \dot{\theta}^2 \\
    \dot{\phi}^2
  \end{bmatrix}
\end{multline}
Multiplying out the matrix equations,

\begin{multline}
    K_\theta =  (A + C \cos{\alpha}) \ddot{\theta} + (B + C
                  \cos{\alpha}) \, \ddot{\phi} \\
    + C \sin{\alpha} \, (-\dot{\theta}^2 +
               \dot{\phi}^2)
\end{multline}

\begin{equation}
    K_\alpha = C \cos{\alpha} \, \ddot{\theta} + B \ddot{\phi} - C
          \sin{\alpha} \, \dot{\theta}^2
\end{equation}
Thus, given ($\theta_0$, $\dot{\theta}_0$, $\ddot{\theta}_0$) and
($\phi_0$, $\dot{\phi}_0$, $\ddot{\phi}_0$), the couples $K_\theta$ and $K_\alpha$
are determined.

\begin{figure}[!htb]
  \begin{center}
    \includegraphics[width=\columnwidth,angle=00]{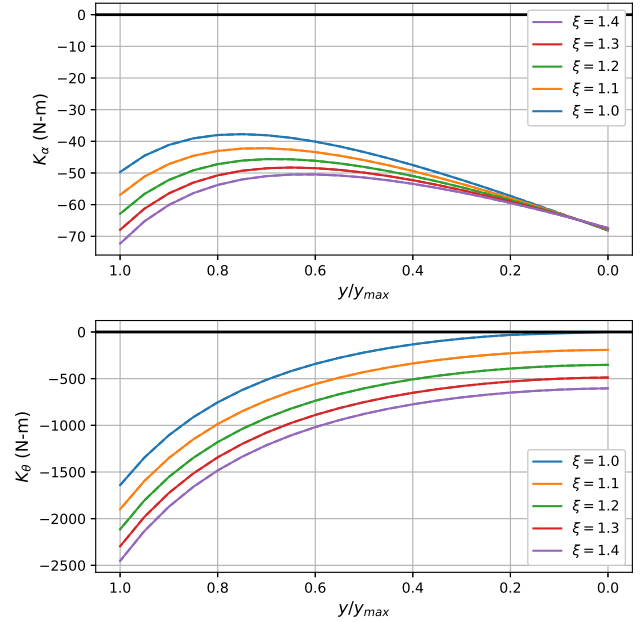}
    \caption{Couples $K_\alpha$ and $K_\theta$ for points along the
      target line with $\xi \ge 1$. $K_\alpha$ is robustly negative,
      of magnitude -\SI{50}{\newton \meter}.  As the radius of
      curvature decreases, the magntude of negative force couple
      $K_\alpha$ gets smaller.}
    \label{fig:dtl_torques_xi_gt1}
  \end{center}
\end{figure}

\begin{figure}[!htb]
  \begin{center}
    \includegraphics[width=\columnwidth,angle=00]{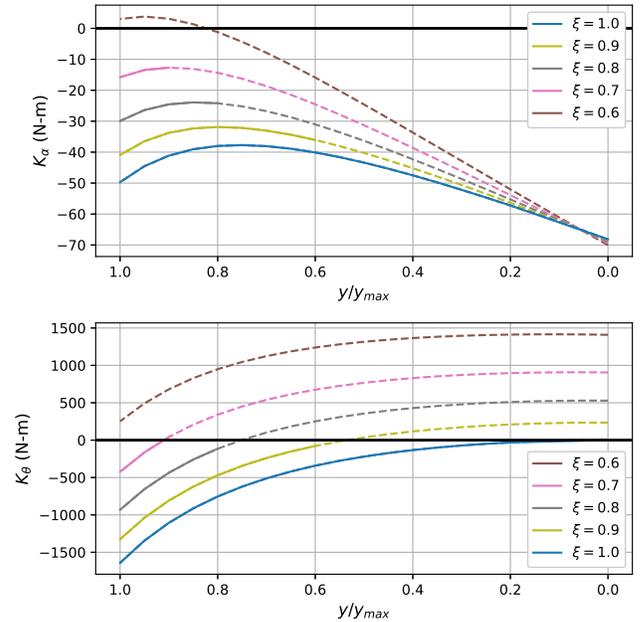}
    \caption{Couples $K_\alpha$ and $K_\theta$ for points along the
      target line with $\xi \le 1$.  As the radius of curvature
      decreases, the magntude of negative force couple $K_\alpha$ gets
      smaller.}
    \label{fig:dtl_torques_xi_lt1}
  \end{center}
\end{figure}

This formalism has been used to calculate the required couples at
points along the target line for various values of $\xi$.  Shown in
Fig.~\ref{fig:dtl_torques_xi_gt1} are results for
$\xi = (1.0, 1.1, 1.2, 1.3, 1.4)$.  The top graphics shows $K_\alpha$
while the lower graphic shows $K_\theta$.  Shown in
Fig.~\ref{fig:dtl_torques_xi_lt1} are results for
$\xi = (0.6, 0.7, 0.8, 0.9, 1.0)$.  The solid lines indicate the
regions where $\ddot{\theta} \le 0$ and $\ddot{\phi} \ge 0$.  The
dashed lines extend beyond this range and are shown for completeness;
however, it is unlikely one would want to implement a solution in
these regions.

$K_\theta$ is the primary couple driving $\ddot{\theta}$, which is
decelerating into impact. Thus, it is not surprising $K_\theta<0$.
The absolute scale of $K_\theta$ depends linearly on our choice of the
inertial moment of the proximal arm of the pendulum.  In these
numerical experiments, that value was chosen by fiat and is not based
on a biomechanical model.  Thus, the absolute scale of $K_\theta$ is
not meaningful.

There is likely some surprise that $K_\alpha$ is negative, as it was
discussed above that $\ddot{\phi}>0$.  The reason $K_\alpha < 0$ is
described in detail in the next section.  The magnitude of $K_\alpha$
in these calculations should be close to what is observed in
experiment, as the inertial properties of the distal arm of the double
pendulum are based on those of a golf club.  Note that in all
realizable cases $K_\alpha$ is negative with magnitude of order tens
of $\SI{}{\newton\meter}$.  This is consistent with the experiments of
MacKenzie \cite{MacKenzie:2016d}, and is the central point of this
paper.

\subsection{Whence art thou, $K_\alpha <  0$ \; (I)}

One can get a sense for why the $K_\alpha < 0$ by considering
Eq.(\ref{eom_phi}), above.  Re-arranging,
\begin{equation}
  \label{eq:club_rotation}
B \ddot{\phi} = - C \ddot{\theta} \cos{\alpha} + C \dot{\theta}^2 \sin{\alpha} + K_\alpha
\end{equation}
This is the equation of motion for the rotation of the club in the
non-inertial frame of the handle of the club (i.e. at the hinge
between the proximal and distal arms of the double pendulum).  The
parameter $B$ is the moment of inertia of the club about the handle.
What follows on the right hand side are the various torques which
drive rotation about the handle, in the frame of reference of the
handle.  Because $\ddot{\phi}>0$, the total torque on the club is
positive.

The first two terms on the left hand side are fictitious forces due to
the fact the position of the handle defines the origin of a
non-inertial reference frame.  The first term is the torque due to the
Euler force associated with the linear acceleration of the handle,
acting through the center of mass of the golf club.  The second term
is the centrifugal force associated with the rotation of the handle
about the hub, acting through the center of mass of the golf club.
The final term is the couple $K_\alpha$.

\begin{figure}[!htb]
  \begin{center}
    \includegraphics[width=\columnwidth,angle=00]{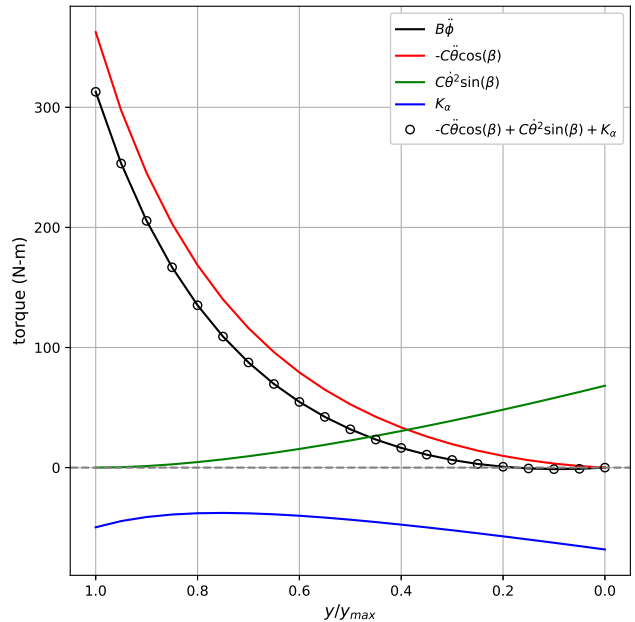}
    \caption{Detail of the terms in the equation of motion for the
      rotation of the club in the frame of reference of the handle of
      the club, for the case $\xi$=1.  The varous curves in graphic
      correspond to terms defined in Eq.(\ref{eq:club_rotation}).  The
      black curve is the torque required to keep the club head moving
      on the specified radius of curvature.  The red and green curves
      indicate the torques associated with the fictitious Euler and
      centrifugal forces due to the acceleration of the non-inertial
      reference frame.  The sum of these two torques is always larger
      than the that of the black curve.  To moderate these two forces,
      a negative couple is applied.  This is shown as the blue curve.
      The sum of the red, blue, and green curves is shown as the black
      open circles, and is equal to the black line.}
    \label{fig:dtl_understanding_torque_I}
  \end{center}
\end{figure}

The four terms in this equation of motion are shown in
Fig.~\ref{fig:dtl_understanding_torque_I} for the case $\xi=1$, for
points along the target line.  The solid black line is
$B \ddot{\phi}$, the solid red line is
$- C \ddot{\theta} \cos{\alpha}$, the solid green line is
$C \dot{\theta}^2 \sin{\alpha}$, the solid blue line is $K_\alpha$,
and the black open circles are the sum of the terms
$C \ddot{\theta} \cos{\alpha}$, $C \dot{\theta}^2 \sin{\alpha}$, and
$K_\alpha$.  Near to $y=0$, the fictitious centrifugal torque
dominates the release (i.e. the green curve), while closer to
$y=y_{max}$ the torque is dominated by the fictitious Euler force
(i.e. the red curve).  In all cases the sum of the Euler and
centrifugal torques are larger than $B \ddot{\phi}$ (i.e. the black
curve).  Thus, to achieve the requisite motion of the club one must
include the couple $K_\alpha < 0$ (i.e. the blue curve).  The sum of
the red, green, and blue curves (i.e. the right hand side of the
equation of motion) is represented as the black open circles,
verifying that they equal the black line.

\section{Calculating the Club Path using Lagrangian Dynamics}

In this section the equations of motion for the double pendulum are
used to calculate the motion of the double pendulum near to impact.
The initial conditions are obtained from the considerations of the
previous sections with $\xi=1$ and $y_0 = 0.5 \, y_{max}$.  The
couples $K_\theta$ and $K_\alpha$ are assumed constant over the range
of motion, and set equal to the values required at impact from the
considerations above.  The value of $K_\alpha$ is \SI{43.4}{\newton
  \meter}.

\begin{figure}[!htb]
  \begin{center}
    \includegraphics[width=\columnwidth,angle=00]{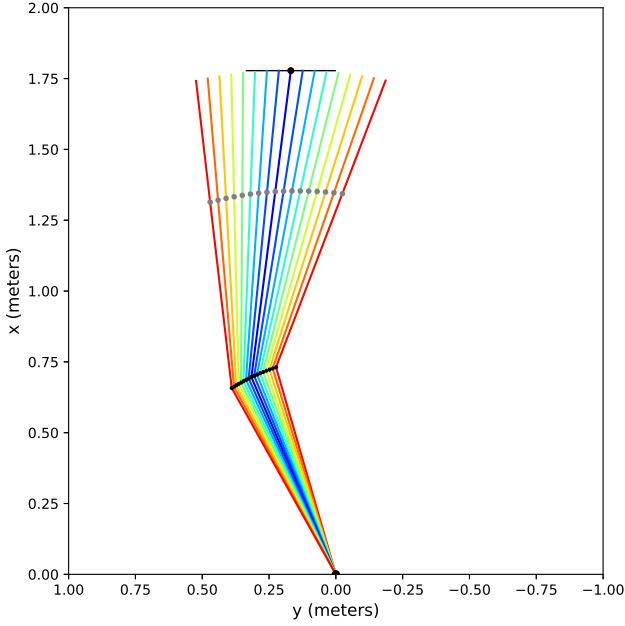}
    \caption{Calculated path of the double pendulum near to impact for
      the parameters $\xi=1$ and $y_0 = 0.5 \, y_{max}$.  The couples
      $K_\theta$ and $K_\alpha$ are assumed constant over the range of
      motion, and chosen to match the initial conditions.}
    \label{fig:dtl_calculated_path}
  \end{center}
\end{figure}

Shown in Fig.~\ref{fig:dtl_calculated_path} is the calculated path of
the double pendulum as it moves through impact.  There is no actual
impact with a golf ball in this calculation, so the club moves
unimpeded through impact.  The gray circles mark the center of mass of
the golf club.  The hinge between the proximal and distal arms is
marked as small black circles.  The point of impact is marked as a
larger black circle on the target line.

\begin{figure}[!htb]
  \begin{center}
    \includegraphics[width=\columnwidth,angle=00]{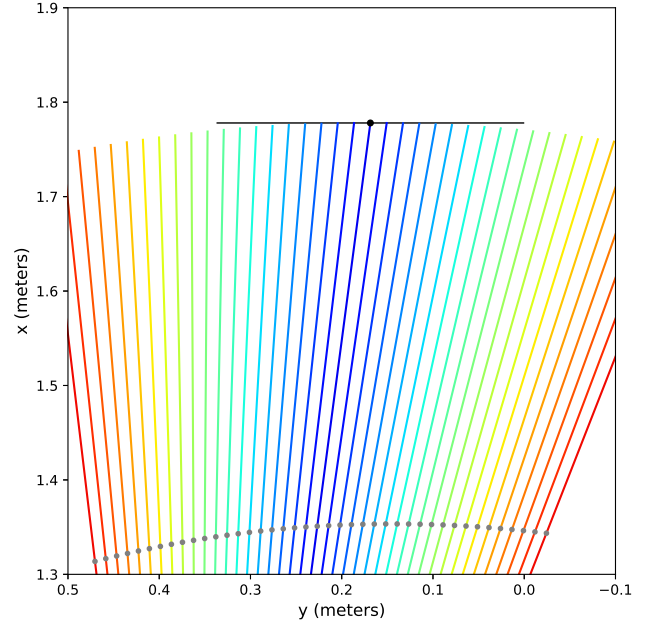}
    \caption{Zoomed in view of the calculated path of the double
      pendulum near to impact.  The time between increments is
      $\SI{0.4}{\milli\second}$ increments, which is of order the
      amount of time the ball stays on the club face during impact.
      The club head remains within fractions of an inch of the target
      line near to impact for time scales longer than that of impact.}
    \label{fig:dtl_calculated_path_zoom}
  \end{center}
\end{figure}

Fig.~\ref{fig:dtl_calculated_path_zoom} zooms in on the target line so as
to show that the distal end of the double pendulum does in fact travel
reasonably straight down the target line at impact, on a path which
has some curvature.  It was verified the radius of curvature equals
$R_1 + R_2 - \delta$, consistent with $\xi = 1$.

The club head is moving at 100 mph (44.7 m/s) at impact.  The
calculation is done with step sizes of $\SI{0.2}{\milli\second}$.
Fig.~\ref{fig:dtl_calculated_path} shows the position of the club in
$\SI{1.0}{\milli\second}$ increments, and the entire simulation covers
a time span of only $\SI{16}{\milli\second}$.
Fig.~\ref{fig:dtl_calculated_path_zoom} shows the position of the club
in $\SI{0.4}{\milli\second}$ increments, which is of order the amount
of time the ball stays on the club.  As can be seen in
Fig.~\ref{fig:dtl_calculated_path_zoom}, the club is moving
approximately straight down the target line on the time scale of
impact.

\begin{figure}[!htb]
  \begin{center}
    \includegraphics[width=\columnwidth,angle=00]{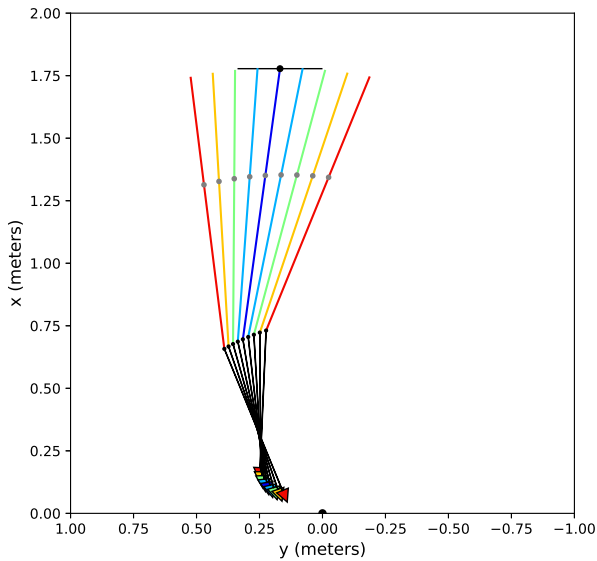}
    \caption{The forces that move the center of mass of the golf club.
      The forces are obtained from the inverse dynamics analysis.  The
      forces are shown as arrows being applied at the handle of the
      club (i.e at the hinge between the proximal and distal arms of
      the double pendulum).  As is shown, they are all oriented in the
      general direction of the hub.  The magnitude of the force at
      impact is $\SI{260}{\newton}$}
    \label{fig:dtl_forces}
  \end{center}
\end{figure}

The solid gray circles indicate the positions of the center of mass of
the golf club.  From these points the inverse dynamics problem can be
solved for the linear forces that move the center of mass.  This
inverse dynamics calculation is meant to enable comparision with the
inverse dynamics anaysis of golf club motion, as implemented by
MacKenzie \cite{MacKenzie:2016d}, Kwon \cite{Kwon:2017a} and Nesbit
\cite{Nesbit:2005b}.  The forces are depicted in
Fig.~\ref{fig:dtl_forces}, shown as arrows acting at the handle.  Note
that they all point in the general direction of the hub, which is
consistent with the measurements of MacKenzie \cite{MacKenzie:2016d}.
The scale is not indicated in the figure, but is of order
$\SI{260}{\newton}$.

The forces obtained using inverse dynamics can be compared with
theory.  The double pendulum imposes constraints on the motion of the
proximal and distal arms, such as the fixed pivot around which the
proximal arm rotates and the hinged connection between two arms.
These constraints result in forces that constrain the motion of the
system, but are not explicit in the Lagrangian.  The implicit force
due to constraints acting on the club can be described as the sum of
four terms.  The first two terms originate from the dynamics of the
proximal arm.  They look as if the center of mass of the distal arm
were located at the hinge,

\begin{equation} \label{eq:F_ddot_theta}
  F_{\ddot{\theta}} =  \bigl( -\sin\theta \, \hat{x} +  \cos\theta \,
  \hat{y} \bigr)
  \, M_2 R_1 \ddot{\theta}
\end{equation}

\begin{equation} \label{eq:F_dot_theta_2}
  F_{\dot{\theta}} =  \bigl( -\cos\theta \, \hat{x} - \sin\theta \, \hat{y}
  \bigr) \, M_2 R_1 \dot{\theta}^2 
\end{equation}
The second two terms involve the dynamics of the distal arm

\begin{equation} \label{eq:F_ddot_phi}
  F_{\ddot{\phi}} =  \bigl( -\sin\phi \, \hat{x} +  \cos\phi \, \hat{y} \bigr)
  \, M_2 <R_2> \ddot{\phi}
\end{equation}

\begin{equation} \label{eq:F_dot_phi_2}
  F_{\dot{\phi}} =  \bigl( -\cos\phi \, \hat{x} - \sin\phi \, \hat{y} \bigr)
  \, M_2 <R_2> \dot{\phi}^2
\end{equation}

\begin{figure}[!htb]
  \begin{center}
    \includegraphics[width=\columnwidth,angle=00]{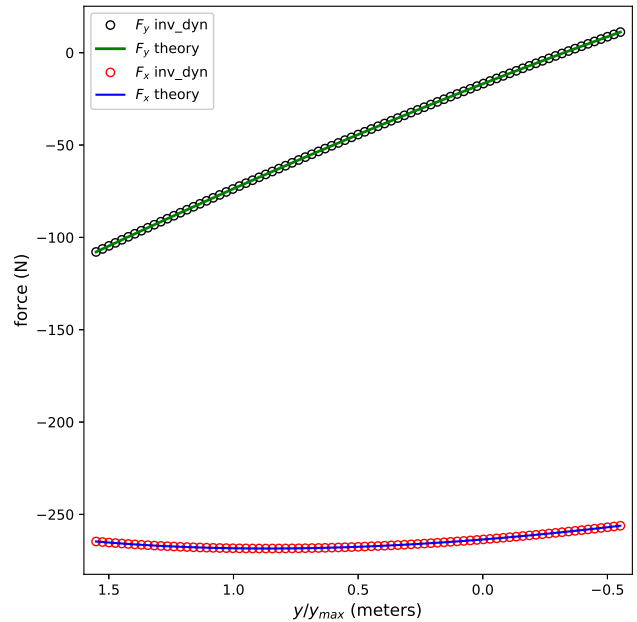}
    \caption{The components $F_x$ and $F_y$ of the forces moving the
      center of mass of the golf club.  The solid lines are calculated
      from theory, as described in the text.  The open circles are
      obtained from the inverse dynamics analysis.  The forces
      obtained from the inverse dynamics analysis are shown to recover
      the forces calculated from theory.}
    \label{fig:dtl_force_check}
  \end{center}
\end{figure}

The sum of the x-components and y-components of these forces are shown
in Fig.~\ref{fig:dtl_force_check} in comparison with the forces
obtained from inverse dynamics.  The solid lines are calculated from
the theoretical expressions, above.  The open circles are obtained
from the inverse dynamics.  Indeed, the inverse dynamics recover the
theoretical answer.

\subsection{Whence art thou, $K_\alpha <  0$ \; (II)}
The results of the previous section allow us to calculate the torques
on the club about the center of mass of the club in the reference
frame of the center of mass of the club.  While the position of the
center of mass defines the origin of a non-inertial reference frame,
the fictitious forces associated with the acceleration of this
reference frame act through the center of mass and thus yield no
torque because the moment arm is zero.  This is why the center of mass
reference frame is always a particularly convenient frame of reference
from which to calculate torques \cite{Feynman_v1_c19:1963}.

There are only two torques which are relevant.  The first is the
torque generated by the linear force that move the center of mass,
detailed in Eqs.~(\ref{eq:F_ddot_theta}) - (\ref{eq:F_dot_phi_2}).
MacKenzie refers to this torque as the moment of force, and is
indicated here as $M_\alpha$.  The other torque is the couple
$K_\alpha$.  Combined, these two torques must equal the total torque
which rotates the club,
$I_{cm} \ddot{\phi} = T_\alpha = M_\alpha + K_\alpha$.

\begin{figure}[!htb]
  \begin{center}
    \includegraphics[width=\columnwidth,angle=00]{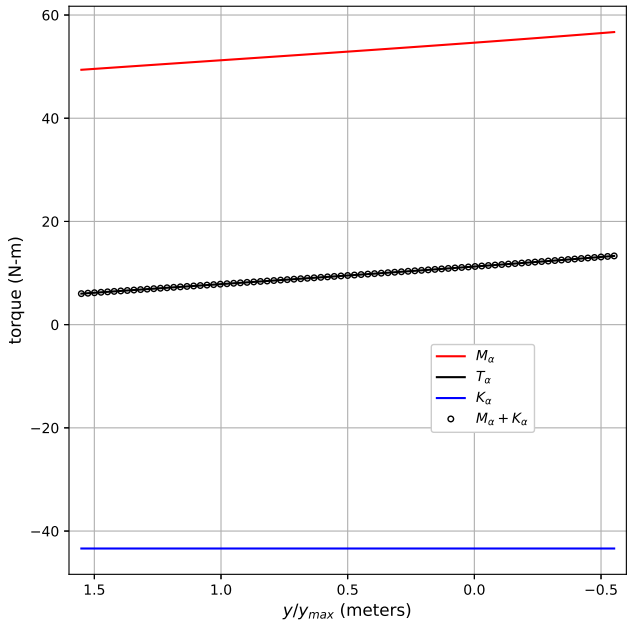}
    \caption{Accounting of the torques in the frame of reference of
      the center of mass of the golf club.  The solid red line is
      $M_\alpha$.  The blue line is $K_\alpha$ = -\SI{43.4}{\newton
        \meter}.  The solid black line is
      $T_\alpha = I_{cm} \ddot{\phi}$.  The open black circles are
      calculated as the sum $M_\alpha + K_\alpha$.  This analysis
      confirms $T_\alpha = M_\alpha + K_\alpha$.  This result is the
      central point to the paper: $M_\alpha$ by itself is larger than
      the requried torque $T_\alpha$.  To compensate for this, the
      torque $K_\alpha<0$ must be applied so as to keep the club
      moving on the path defined by the radius of curvature.}
    \label{fig:dtl_torques_in_cm_frame}
  \end{center}
\end{figure}

These torques are shown in Fig.~\ref{fig:dtl_torques_in_cm_frame} as
the club moves through impact, from the simulation above.  The solid
red line is $M_\alpha$.  The blue line is $K_\alpha$ =
\SI{43.4}{\newton \meter}.  The solid black line is
$T_\alpha = I_{cm} \ddot{\phi}$.  The open black circles are
calculated as the sum $M_\alpha + K_\alpha$.  This analysis confirms
$T_\alpha = M_\alpha + K_\alpha$.

Once again, we see that while the total torque on the club
$T_\alpha>0$, the couple $K_\alpha$ has to be negative because the
other torque in the problem $M_\alpha$ would otherwise provide more
torque than what is required to move the club head on the path defined
by the radius of curvature.

It is interesting to point out that the value for $K_\alpha$ was set
by balancing torques in the non-inertial frame of reference of the
handle of the club.  In this section the analysis was done in the
non-inertial frame of reference of the center of mass of the club.  In
both cases, the couple $K_\alpha$ has the same value.  This serves to
emphasize that if you solve for the forces and torques which move a
rigid body in multiple reference frames, even non-inertial reference
frames, you should always recover the same answer.

\section{Seeking A  Match to MacKenzie's Data}
The majority of MacKenzie's video `In-Plane Couple and Moment of Force
During the Golf Swing' \cite{MacKenzie:2016d} highlights the golf
swing of a single golfer.  For this golfer in the last frame before
impact, the club head is moving at 116.5 mph and the measured values
of force and torques are force $F_0$ = \SI{456}{\newton}, moment of
force $M_0$ = \SI{55.8}{\newton\meter}, and couple $K_0$ =
\SI{-59.1}{\newton\meter}.  In this section the double pendulum model
is solved at impact over a grid of parameter values $\delta$, $\xi$,
and $y_0$, in an attempt to find the best fit to $F_0$, $M_0$ and
$K_0$.  All other parameters in the problem, such as the length of
arms of the double pendulum and the inertial properties of the golf
club, are as defined in Appendix~\ref{appdx:ap1}.  As such they are
just approximations to what may have been used in the experiments of
MacKenzie.

\begin{figure}[!htb]
  \begin{center}
    \includegraphics[width=\columnwidth,angle=00]{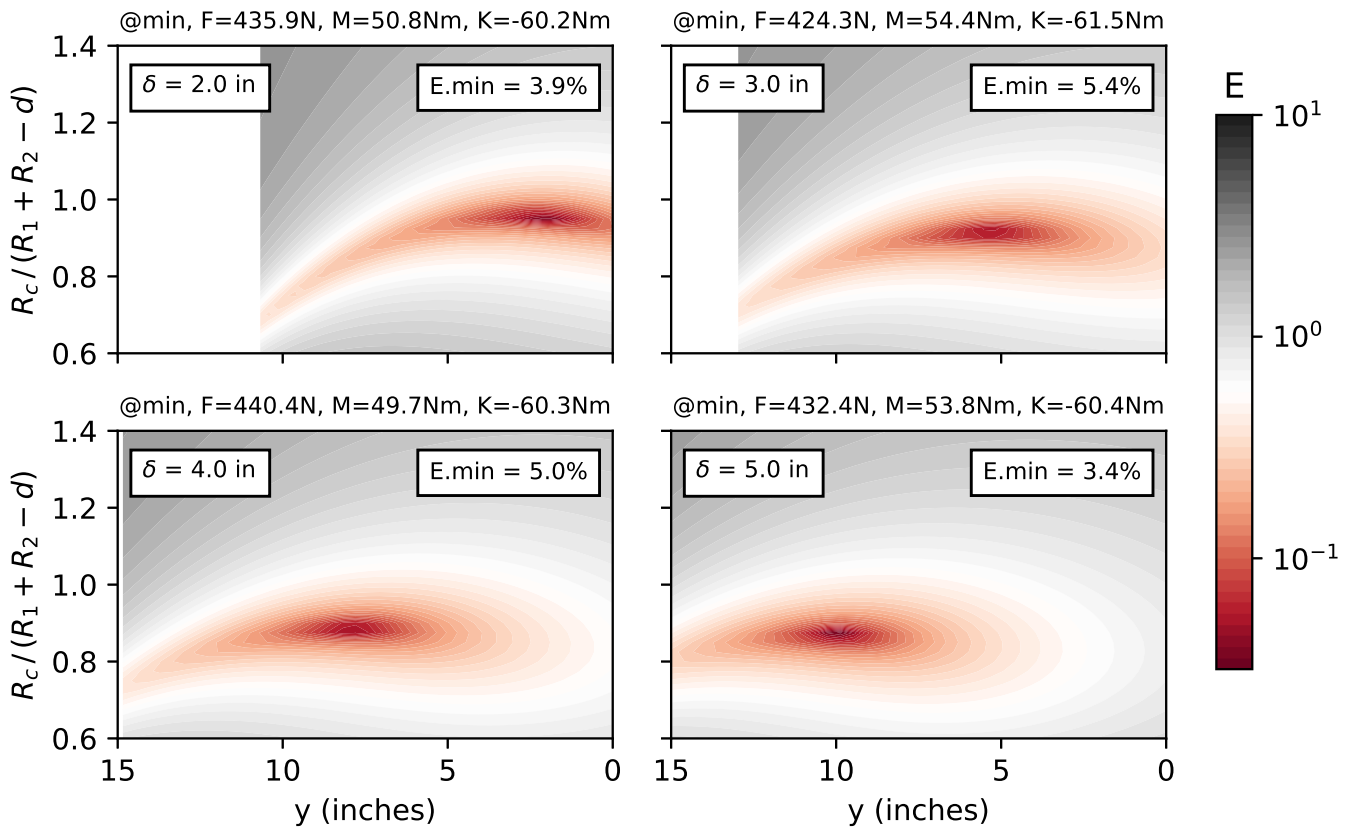}
    \caption{The results of a search over a grid of parameters in an
      attempt to match to force $F$ and torques, $M$ and $K$, reported
      by MacKenzie.  The the color scale encodes $\log_{10}(E)$, where
      $E$ is the average fractional error, as described in the text.
      The value of $\delta$ is given in the top left corner of each
      panel.  The minimum error is indicated in the top right corner
      of each panel.  The values $F$, $M$, and $K$ at the minimum are
      listed at the top of each panel.  The values to which they are
      being fit are $F_0$ = \SI{456}{\newton}, $M_0$ =
      \SI{55.8}{\newton\meter}, and $K_0$ = \SI{-59.1}{\newton\meter}.
      It is a primary result of this paper that the double pendulum
      model of the golf swing is able to obtain the results reported
      by MacKenzie to within a few percent. }
    \label{fig:force_torque_matcher}
  \end{center}
\end{figure}

The result of this search is summarized in the charts of
Fig.~\ref{fig:force_torque_matcher}.  The different panels correspond
to different values of $\delta$, ranging from 2 to 5 inches.  The
abscissa corresponds to different values of $y_0$, ranging from 0 to
15 inches.  The ordinate corresponds to different values of $\xi$ in
the range 0.6-1.4.  The color scale encodes the difference between
MacKenzie's data ($F_0$, $M_0$, and $K_0$) and the result obtained
from the model ($F$, $M$, and $K$).  This difference is calculated as
the sum-of-squares average fractional error $E$,

\begin{equation}
E^2 = \frac{1}{3} \bigl( (\frac{F-F_0}{F_0})^2 + (\frac{m -
  m_0}{m_0})^2 + (\frac{K - K_0}{K_0})^2 \bigr).  
\end{equation}
This treats the parameters $F, M$, and $K$ as if they were
independent.  To this end, $m = M/F$, and is thus only sensitive to
the angle between $F$ and the shaft of the club.  To accommodate the
dynamic range, the color scale encodes $\log_{10}(E)$.

The value of $\delta$ is given in the top left corner of each panel.
The minimum value of $E$ is indicated in the top right corner of each
panel.  The values $F$, $M$, and $K$ at the minimum are listed at the
top of each panel.  It is a primary result of this paper that the
double pendulum model of the golf swing is able to obtain the force
and torques reported by MacKenzie to within a few percent.  It is
possible the scale of these differences are consistent with the
instrumental noise in MacKenzie's experiments.

These data show that one can use the same set of forces and torques
$F$, $M$, and $K$ to hit the ball standing different distances from
the ball, $\delta$, and from different positions in the stance, $y_0$.
As the ball is moved further forward in the stance, the golfer must
stand closer to the ball and the radius of curvature of the club head
path becomes smaller.

\section{Speculation About How $K_\alpha$  is Generated}
The scale of $K_{\alpha}$ is of order 50 N-m.  What can generate a
couple of this magnitude?  This section explores three possibilities.

It is important to remember this particular torque is a force couple.  It
can be thought of as being generated by two linear force vectors,
equal in magnitude $F_K$ but opposite in direction, separated through
a distance $d$.  Because the linear sum of the forces is zero, there
is no net force on the center of mass of the object due to the two
force vectors.  However, because they are separated through the
distance $d$, they yield a torque of magnitude $d \; F_K$
perpendicular to the plane defined by the two force vectors, and thus
generate rotation.

\subsection{The Hands}
Suppose this couple is generated by forces applied by the hands.  This
could be either because the hands are actively applying force, or
because the hands can not keep up with the linear and/or rotational
speeds at impact.

For a right handed golfer, imagine the left hand applying a force in
the direction of motion of the club, and the right hand applying a
force of equal magnitude in the opposite direction (i.e. opposing the
motion of the club).  Suppose the distance from the pinky finger of
the left hand to the forefinger of the right hand when a right handed
golfer grips the club is 1/6 meter (i.e. 6-7 inches) and is the
distance through which the couple is applied.  Then to generate a
couple applying $\SI{50}{\newton\meter}$ of torque, each hand would
have to be applying $\SI{300}{\newton}$ of force in opposite
directions.  This is in addition to the hundreds of Newtons of linear
force already discussed above, which is presumably split between the
two hands.  $\SI{300}{\newton}$ of force amounts to 70 lbs force.
That seems like a lot of force for each hand to be applying.  For this
reason, it would seem that this explanation alone is insufficient to
provide all of $K_\alpha$.

However, it is important to note that this negative couple is applied
only 10-20 ms before impact.  Thus, the resulting impulse (i.e. torque
multiplied by time) is not particularly large.  If this torque were
due to the fact the hands can not keep up with the release of the
club, it might be difficult for the golfer to perceive this applied
torque.  It would be quite spectacular if golfers have learned to
harness this natural drag to help them to hit the ball straighter.

\onecolumngrid 

\twocolumngrid

\subsection{Aerodynamic Drag of the Club head}
Another possible source of negative couple is the aerodynamic drag on
the club head as it approaches impact.  Imagine the size of that force
is $F_d$ in the direction opposing the motion of the club head.  Now
imagine that the hands apply a force of equal magnitude but in the
opposite direction, counter acting this drag.  The separation between
these two forces is the length of the club, $\epsilon = R_2$.
Henrikson reports \cite{Henrikson:2014a} the scale of the drag force
to be 4.5 - 7.5 N.  If we use $\SI{10}{\newton}$ as an upper limit,
and assume a club of length $\SI{1}{\meter}$, then this can yield a
couple of order $\SI{10}{\newton\meter}$.  Again, this is too small to
give values as large as $\SI{50}{\newton\meter}$.

\subsection{Inertia of the squaring of the club face}
Missing from the model of the double pendulum is the fact that the
club face goes from open to square to closed as the club moves through
impact.  This requires rotation of the club around the long axis of
the shaft.  It also requires the rotation of the arms and hands, which
support the club.  This motion is related to the $\beta$-torques and
$\gamma$-torques described in Nesbit's 2005 paper
\cite{Nesbit:2005b}, which involve motion out of the swing plane and
about the axis of the shaft, respectively.

For our purposes, consider that the motion caused by the
$\beta$-torque and $\gamma$-torque is coupled to the release of the
hands, defined in this paper as the angle $\alpha = \theta - \phi$.
It is certainly the case that the club face is open when
$\alpha \approx \pi/2$, it is square near to impact where
$\alpha \approx 0$, and closed after impact, when $\alpha$ ends at
$-\pi/2$.

Now posit that the $\beta$-torques and $\gamma$-torques causes motion
that affects the moment of inertia relevant to the motion in the plane
of the golf swing.  This could involve the relative positions of the arms
and hands, the rotation of the club around its axis, motion of mass
above and below the swing plane, etc.  

Further make the generalization that the kinetic energy associated
with the squaring of the club manifests itself in the swing plane as
$KE_s$ and that this can be parameterized in terms of the angular
speed $\dot{\alpha}$ and a moment of inertia $I_s$,

\begin{equation}
KE_s = \frac{1}{2} I_s \dot{\alpha}^2
\end{equation}
As long as we are only solving the double pendulum in the vicinity of
impact, this additional term can then be included in the Lagrangian of
this paper (i.e. not making generalizations beyond the
immediate vicinity of impact).

With this addition, the equations of motion become 

\begin{equation}
  A \ddot{\theta} + C \ddot{\phi}
  \cos{(\theta - \phi)} + C \dot{\phi}^2 \sin{(\theta -
    \phi)} = K_\theta - K_\alpha + I_s (\ddot{\phi} - \ddot{\theta})
\end{equation}

\begin{equation}
B \ddot{\phi} + C \ddot{\theta}
\cos{(\theta - \phi)} - C \dot{\theta}^2 \sin{(\theta -
  \phi)} = K_\alpha - I_s (\ddot{\phi} - \ddot{\theta})
\end{equation}
As has been shown above, $\ddot{\phi}>0$ and $\ddot{\theta}<0$, so the
term $- I_s (\ddot{\phi} - \ddot{\theta})$ functions as a negative
torque.

In the exercises above, $K_\alpha$ was assumed to provide the full
negative couple required to keep the club moving straight down the
line.  For arguments sake, lets assume here that all of the negative
couple comes from $I_s$.  Evaluating the example above at impact,
$(\ddot{\phi} - \ddot{\theta}) \approx
\SI{500}{\radian\per\square\second}$ which suggests
$I_s \approx \SI{0.1}{\kilogram \square\meter}$.  We can compare this
with the value of the moment of inertial of the golf club about its
handle, $I_{R_2} = \SI{0.24}{\kilogram \square\meter}$ used in this
paper.  Thus, $I_s$ needs to be of order 40\% of the size of
$I_{R_2}$, which would be a large perturbation.  While this seems like
a logical avenue for the biomechanics community to explore, it is
possible it will not be large enough to explain all of $K_\alpha$.

\subsection{Speculation Summary}
This section has explored three physical processes that could generate
$K_\alpha \approx -\SI{50}{\newton\meter}$.  Each one of them
individually seems too small to provide a torque of sufficient
magnitude.  Thus, instead of there being one clean source of
$K_\alpha$, it seems likely the actual answer involves multiple terms,
or phenomena not considered in this paper.

\section{Summary}
Motivated by MacKenzie's observation of a negative couple near to
impact \cite{MacKenzie:2020a, MacKenzie:2016d}, this paper has
explored a model for how the golf club moves near to impact.  It
assumes the club is moving as the distal arm of a double pendulum and
that at impact the club head is moving straight down the target line,
at its maximum speed, on a path of defined curvature.  From this model,
the forces and torques required to move the club near to impact are
calculated.

The results obtained from this model are shown to be quantitatively
consistent with data reported by Mackenzie to within a few percent.
Indeed, the negative couple near to impact is found to be a robust
feature of this model.  It balances torques resulting from the forces
that drive the center of mass of the golf club.  These torques reduce
the radius of curvature of the path of the club head as it moves
through impact.  By applying a negative couple the golfer is able to
achieve a larger radius of curvature.  This reduces the difference
between the path of the club head and the target line as the club head
moves near to impact. Because the negative couple can also serve to
reduce the rotational speed of the club, its presence in the golf
swing manifests a trade between distance and direction.

\appendix
\section{Model Parameters}
\label{appdx:ap1}

The properties of the golf club were taken from Nesbit
\cite{Nesbit:2009a}, for consistency.  They are:

\begin{itemize}
  
\item $R_2 = \SI{1.092}{\meter}$, the length of the golf club in meters.
  Presumably measured from a place between the two hands to the middle
  of the club face.

\item $M_2 = \SI{0.382}{\kilogram}$, the mass of the golf club.
\item $<R_2> = \SI{0.661}{\meter}$, the first moment, which is the
  distance from the hands to the center of mass of the club.  

\item $ I_{2,CM} = M_2 <(R_2 - <R_2>)^2>  =
  \SI{0.071}{\kilogram\square\meter}$, the moment of inertia of the golf club measured about it's center of mass.
\end{itemize}

The properties of the proximal arm of the double pendulum were picked
by fiat, and are not based on any biomechanical model.

\begin{itemize}
\item $R_1 = 0.7 R_2$, the length of the proximal arm of the double
  pendulum.  This number is not based on any detailed measurement.  It
  is meant to be a very crude approximation.
  
\item $ M_1 <(R_1 - <R_1>)^2> = 3 * I_{2, CM}$, the moment of inertia
  of the proximal arm of the double pendulum about the fixed hub.
  This number is just a stab in the dark.  Its only relevance is to
  scale the magnitude of $K_\theta$.
\end{itemize}
  
The distance $\delta$ is taken to be $\SI{7.84}{\cm}$, which is just
about 3.1 inches.  This was chosen so that $R_1 + R_2 - \delta$ = 70
in.  Again, there is no particular reason for this choice other than
it made the length of the accessible points along the target line of
order 12 in.

\newpage

\bibliography{}  

\begin{thebibliography}{22}%
\makeatletter
\providecommand \@ifxundefined [1]{%
 \@ifx{#1\undefined}
}%
\providecommand \@ifnum [1]{%
 \ifnum #1\expandafter \@firstoftwo
 \else \expandafter \@secondoftwo
 \fi
}%
\providecommand \@ifx [1]{%
 \ifx #1\expandafter \@firstoftwo
 \else \expandafter \@secondoftwo
 \fi
}%
\providecommand \natexlab [1]{#1}%
\providecommand \enquote  [1]{``#1''}%
\providecommand \bibnamefont  [1]{#1}%
\providecommand \bibfnamefont [1]{#1}%
\providecommand \citenamefont [1]{#1}%
\providecommand \href@noop [0]{\@secondoftwo}%
\providecommand \href [0]{\begingroup \@sanitize@url \@href}%
\providecommand \@href[1]{\@@startlink{#1}\@@href}%
\providecommand \@@href[1]{\endgroup#1\@@endlink}%
\providecommand \@sanitize@url [0]{\catcode `\\12\catcode `\$12\catcode
  `\&12\catcode `\#12\catcode `\^12\catcode `\_12\catcode `\%12\relax}%
\providecommand \@@startlink[1]{}%
\providecommand \@@endlink[0]{}%
\providecommand \url  [0]{\begingroup\@sanitize@url \@url }%
\providecommand \@url [1]{\endgroup\@href {#1}{\urlprefix }}%
\providecommand \urlprefix  [0]{URL }%
\providecommand \Eprint [0]{\href }%
\providecommand \doibase [0]{http://dx.doi.org/}%
\providecommand \selectlanguage [0]{\@gobble}%
\providecommand \bibinfo  [0]{\@secondoftwo}%
\providecommand \bibfield  [0]{\@secondoftwo}%
\providecommand \translation [1]{[#1]}%
\providecommand \BibitemOpen [0]{}%
\providecommand \bibitemStop [0]{}%
\providecommand \bibitemNoStop [0]{.\EOS\space}%
\providecommand \EOS [0]{\spacefactor3000\relax}%
\providecommand \BibitemShut  [1]{\csname bibitem#1\endcsname}%
\let\auto@bib@innerbib\@empty
\bibitem [{\citenamefont {MacKenzie}\ \emph {et~al.}(2020)\citenamefont
  {MacKenzie}, \citenamefont {McCourt},\ and\ \citenamefont
  {Champoux}}]{MacKenzie:2020a}%
  \BibitemOpen
  \bibfield  {author} {\bibinfo {author} {\bibfnamefont {S.}~\bibnamefont
  {MacKenzie}}, \bibinfo {author} {\bibfnamefont {M.}~\bibnamefont {McCourt}},
  \ and\ \bibinfo {author} {\bibfnamefont {L.}~\bibnamefont {Champoux}},\
  }\href@noop {} {\bibfield  {journal} {\bibinfo  {journal} {International
  Journal of Golf Science}\ }\textbf {\bibinfo {volume} {8}} (\bibinfo {year}
  {2020})}\BibitemShut {NoStop}%
\bibitem [{\citenamefont {MacKenzie}(2016{\natexlab{a}})}]{MacKenzie:2016d}%
  \BibitemOpen
  \bibfield  {author} {\bibinfo {author} {\bibfnamefont {S.}~\bibnamefont
  {MacKenzie}},\ }\href@noop {} {\bibfield  {journal} {\bibinfo  {journal}
  {Vimeo (https://vimeo.com/158856998)}\ } (\bibinfo {year}
  {2016}{\natexlab{a}})}\BibitemShut {NoStop}%
\bibitem [{\citenamefont {Cheethan}\ \emph {et~al.}(2008)\citenamefont
  {Cheethan}, \citenamefont {Rose}, \citenamefont {Hinrichs}, \citenamefont
  {Neal}, \citenamefont {Mottram}, \citenamefont {Hurrion},\ and\ \citenamefont
  {Vint}}]{Cheetham:2008a}%
  \BibitemOpen
  \bibfield  {author} {\bibinfo {author} {\bibfnamefont {P.}~\bibnamefont
  {Cheethan}}, \bibinfo {author} {\bibfnamefont {G.}~\bibnamefont {Rose}},
  \bibinfo {author} {\bibfnamefont {R.}~\bibnamefont {Hinrichs}}, \bibinfo
  {author} {\bibfnamefont {R.}~\bibnamefont {Neal}}, \bibinfo {author}
  {\bibfnamefont {R.}~\bibnamefont {Mottram}}, \bibinfo {author} {\bibfnamefont
  {P.}~\bibnamefont {Hurrion}}, \ and\ \bibinfo {author} {\bibfnamefont
  {P.}~\bibnamefont {Vint}},\ }in\ \href@noop {} {\emph {\bibinfo {booktitle}
  {Science and Golf V: Proceedings of the World Scientific Congress of
  Golf}}},\ \bibinfo {editor} {edited by\ \bibinfo {editor} {\bibfnamefont
  {D.}~\bibnamefont {Crews}}\ and\ \bibinfo {editor} {\bibfnamefont
  {R.}~\bibnamefont {Lutz}}}\ (\bibinfo  {publisher} {Energy In Motion},\
  \bibinfo {address} {Mesa, Arizona},\ \bibinfo {year} {2008})\ pp.\ \bibinfo
  {pages} {30--36}\BibitemShut {NoStop}%
\bibitem [{\citenamefont {{Wikipedia contributors}}(2020)}]{wiki:ForceCouple}%
  \BibitemOpen
  \bibfield  {author} {\bibinfo {author} {\bibnamefont {{Wikipedia
  contributors}}},\ }\href@noop {} {\bibfield  {journal} {\bibinfo  {journal}
  {Wikipedia, The Free Encyclopedia}\ } (\bibinfo {year} {2020})}\BibitemShut
  {NoStop}%
\bibitem [{\citenamefont {Fishman}(1978)}]{Fishman:1978b}%
  \BibitemOpen
  \bibfield  {author} {\bibinfo {author} {\bibfnamefont {L.}~\bibnamefont
  {Fishman}},\ }\href@noop {} {\enquote {\bibinfo {title} {A hot, new kind of
  arm swing},}\ }\bibinfo {howpublished} {Golf Magazine, 9/78, pp. 54-57.}
  (\bibinfo {year} {1978})\BibitemShut {NoStop}%
\bibitem [{\citenamefont {Malaska}(2018)}]{Malaska:2018a}%
  \BibitemOpen
  \bibfield  {author} {\bibinfo {author} {\bibfnamefont {M.}~\bibnamefont
  {Malaska}},\ }\href@noop {} {\enquote {\bibinfo {title} {The rotor drill -
  joe nichols' favorite golf swing drill},}\ }\bibinfo {howpublished}
  {https://www.youtube.com/watch?v=xohaFjlKDtk} (\bibinfo {year}
  {2018})\BibitemShut {NoStop}%
\bibitem [{\citenamefont {MacKenzie}(2016{\natexlab{b}})}]{MacKenzie:2016a}%
  \BibitemOpen
  \bibfield  {author} {\bibinfo {author} {\bibfnamefont {S.}~\bibnamefont
  {MacKenzie}},\ }\href@noop {} {\bibfield  {journal} {\bibinfo  {journal}
  {Vimeo (https://vimeo.com/158419250)}\ } (\bibinfo {year}
  {2016}{\natexlab{b}})}\BibitemShut {NoStop}%
\bibitem [{\citenamefont {MacKenzie}(2016{\natexlab{c}})}]{MacKenzie:2016b}%
  \BibitemOpen
  \bibfield  {author} {\bibinfo {author} {\bibfnamefont {S.}~\bibnamefont
  {MacKenzie}},\ }\href@noop {} {\bibfield  {journal} {\bibinfo  {journal}
  {Vimeo (https://vimeo.com/162015461)}\ } (\bibinfo {year}
  {2016}{\natexlab{c}})}\BibitemShut {NoStop}%
\bibitem [{\citenamefont {MacKenzie}(2016{\natexlab{d}})}]{MacKenzie:2016c}%
  \BibitemOpen
  \bibfield  {author} {\bibinfo {author} {\bibfnamefont {S.}~\bibnamefont
  {MacKenzie}},\ }\href@noop {} {\bibfield  {journal} {\bibinfo  {journal}
  {Vimeo (https://vimeo.com/160385937)}\ } (\bibinfo {year}
  {2016}{\natexlab{d}})}\BibitemShut {NoStop}%
\bibitem [{\citenamefont {Kwon}(2017)}]{Kwon:2017a}%
  \BibitemOpen
  \bibfield  {author} {\bibinfo {author} {\bibfnamefont {Y.-H.}\ \bibnamefont
  {Kwon}},\ }\href@noop {} {\enquote {\bibinfo {title} {Hand-club interaction:
  1. inverse dynamics},}\ }\bibinfo {howpublished}
  {http://drkwongolf.info/technotes/mh\_kinetics.pdf} (\bibinfo {year}
  {2017})\BibitemShut {NoStop}%
\bibitem [{\citenamefont {Nesbit}\ and\ \citenamefont
  {Serrano}(2005)}]{Nesbit:2005b}%
  \BibitemOpen
  \bibfield  {author} {\bibinfo {author} {\bibfnamefont {S.~M.}\ \bibnamefont
  {Nesbit}}\ and\ \bibinfo {author} {\bibfnamefont {M.}~\bibnamefont
  {Serrano}},\ }\href@noop {} {\bibfield  {journal} {\bibinfo  {journal}
  {Journal of Sports Science and Medicine}\ }\textbf {\bibinfo {volume} {4}},\
  \bibinfo {pages} {520} (\bibinfo {year} {2005})}\BibitemShut {NoStop}%
\bibitem [{\citenamefont {Nesbit}\ and\ \citenamefont
  {McGinnis}(2009)}]{Nesbit:2009a}%
  \BibitemOpen
  \bibfield  {author} {\bibinfo {author} {\bibfnamefont {S.~M.}\ \bibnamefont
  {Nesbit}}\ and\ \bibinfo {author} {\bibfnamefont {R.~S.}\ \bibnamefont
  {McGinnis}},\ }\href@noop {} {\bibfield  {journal} {\bibinfo  {journal}
  {Journal of Sports Science and Medicine}\ }\textbf {\bibinfo {volume} {8}},\
  \bibinfo {pages} {235} (\bibinfo {year} {2009})}\BibitemShut {NoStop}%
\bibitem [{\citenamefont {Nesbit}\ and\ \citenamefont
  {McGinnis}(2014)}]{Nesbit:2014a}%
  \BibitemOpen
  \bibfield  {author} {\bibinfo {author} {\bibfnamefont {S.~M.}\ \bibnamefont
  {Nesbit}}\ and\ \bibinfo {author} {\bibfnamefont {R.~S.}\ \bibnamefont
  {McGinnis}},\ }\href@noop {} {\bibfield  {journal} {\bibinfo  {journal}
  {Journal of Sports Science and Medicine}\ }\textbf {\bibinfo {volume} {13}},\
  \bibinfo {pages} {859} (\bibinfo {year} {2014})}\BibitemShut {NoStop}%
\bibitem [{\citenamefont {Cochran}\ and\ \citenamefont
  {Stobbs}(1968)}]{Cochran:1968}%
  \BibitemOpen
  \bibfield  {author} {\bibinfo {author} {\bibfnamefont {A.}~\bibnamefont
  {Cochran}}\ and\ \bibinfo {author} {\bibfnamefont {J.}~\bibnamefont
  {Stobbs}},\ }\href@noop {} {\emph {\bibinfo {title} {Search for the Perfect
  Swing}}}\ (\bibinfo  {publisher} {The Golf Society of Great Britan},\
  \bibinfo {year} {1968})\BibitemShut {NoStop}%
\bibitem [{\citenamefont {Jorgensen}(1970)}]{Jorgensen:1970}%
  \BibitemOpen
  \bibfield  {author} {\bibinfo {author} {\bibfnamefont {T.}~\bibnamefont
  {Jorgensen}},\ }\href@noop {} {\bibfield  {journal} {\bibinfo  {journal}
  {American Journal of Physics}\ }\textbf {\bibinfo {volume} {38}},\ \bibinfo
  {pages} {644} (\bibinfo {year} {1970})}\BibitemShut {NoStop}%
\bibitem [{\citenamefont {Jorgensen}(1994)}]{Jorgensen:1994a}%
  \BibitemOpen
  \bibfield  {author} {\bibinfo {author} {\bibfnamefont {T.}~\bibnamefont
  {Jorgensen}},\ }\href@noop {} {\emph {\bibinfo {title} {The Physics of
  Golf}}}\ (\bibinfo  {publisher} {American Institute of Physics},\ \bibinfo
  {address} {New York},\ \bibinfo {year} {1994})\BibitemShut {NoStop}%
\bibitem [{\citenamefont {Nesbit}\ and\ \citenamefont
  {Jacobs}(2019)}]{Nesbit:2019_MR}%
  \BibitemOpen
  \bibfield  {author} {\bibinfo {author} {\bibfnamefont {S.}~\bibnamefont
  {Nesbit}}\ and\ \bibinfo {author} {\bibfnamefont {M.}~\bibnamefont
  {Jacobs}},\ }\href@noop {} {\enquote {\bibinfo {title} {Dr. nesbit and
  michael jacobs discuss the science of the golf swing: Motional resistance
  (35:42)},}\ }\bibinfo {howpublished}
  {https://www.youtube.com/watch?v=KbdDHIJqGgA} (\bibinfo {year}
  {2019})\BibitemShut {NoStop}%
\bibitem [{\citenamefont {Anton}\ \emph {et~al.}(2009)\citenamefont {Anton},
  \citenamefont {Bivens}, \citenamefont {Davis},\ and\ \citenamefont
  {Polaski}}]{Anton:2009a}%
  \BibitemOpen
  \bibfield  {author} {\bibinfo {author} {\bibfnamefont {H.}~\bibnamefont
  {Anton}}, \bibinfo {author} {\bibfnamefont {I.}~\bibnamefont {Bivens}},
  \bibinfo {author} {\bibfnamefont {S.}~\bibnamefont {Davis}}, \ and\ \bibinfo
  {author} {\bibfnamefont {T.}~\bibnamefont {Polaski}},\ }\href@noop {} {\emph
  {\bibinfo {title} {Calculus Multivariable}}},\ \bibinfo {edition} {9th}\ ed.\
  (\bibinfo  {publisher} {John Wiley \& Sons, Inc.},\ \bibinfo {address}
  {Hoboken, NY 07030},\ \bibinfo {year} {2009})\ Chap.~\bibinfo {chapter}
  {12}\BibitemShut {NoStop}%
\bibitem [{\citenamefont {Goldstein}(1981)}]{Goldstein:1981}%
  \BibitemOpen
  \bibfield  {author} {\bibinfo {author} {\bibfnamefont {H.}~\bibnamefont
  {Goldstein}},\ }\href@noop {} {\emph {\bibinfo {title} {Classical
  Mechanics}}},\ \bibinfo {edition} {2nd}\ ed.\ (\bibinfo  {publisher}
  {Addison-Wesley},\ \bibinfo {address} {Reading, Ma.},\ \bibinfo {year}
  {1981})\BibitemShut {NoStop}%
\bibitem [{\citenamefont {Symon}(1971)}]{Simon:1971}%
  \BibitemOpen
  \bibfield  {author} {\bibinfo {author} {\bibfnamefont {K.}~\bibnamefont
  {Symon}},\ }\href@noop {} {\emph {\bibinfo {title} {Mechanics}}},\ \bibinfo
  {edition} {3rd}\ ed.\ (\bibinfo  {publisher} {Addison-Wesley},\ \bibinfo
  {address} {Reading, Ma.},\ \bibinfo {year} {1971})\BibitemShut {NoStop}%
\bibitem [{\citenamefont {Feynman}\ \emph {et~al.}(1977)\citenamefont
  {Feynman}, \citenamefont {Leighton},\ and\ \citenamefont
  {Sands}}]{Feynman_v1_c19:1963}%
  \BibitemOpen
  \bibfield  {author} {\bibinfo {author} {\bibfnamefont {R.}~\bibnamefont
  {Feynman}}, \bibinfo {author} {\bibfnamefont {R.}~\bibnamefont {Leighton}}, \
  and\ \bibinfo {author} {\bibfnamefont {M.}~\bibnamefont {Sands}},\
  }\href@noop {} {\emph {\bibinfo {title} {The Feynman Lectures on Physics}}},\
  Vol.~\bibinfo {volume} {I}\ (\bibinfo  {publisher} {Addison-Wesley},\
  \bibinfo {address} {Reading, Ma.},\ \bibinfo {year} {1977})\ Chap.~\bibinfo
  {chapter} {19}\BibitemShut {NoStop}%
\bibitem [{\citenamefont {Henrikson}\ \emph {et~al.}(2014)\citenamefont
  {Henrikson}, \citenamefont {Wood},\ and\ \citenamefont
  {Hart}}]{Henrikson:2014a}%
  \BibitemOpen
  \bibfield  {author} {\bibinfo {author} {\bibfnamefont {E.}~\bibnamefont
  {Henrikson}}, \bibinfo {author} {\bibfnamefont {P.}~\bibnamefont {Wood}}, \
  and\ \bibinfo {author} {\bibfnamefont {J.}~\bibnamefont {Hart}},\ }\href@noop
  {} {\bibfield  {journal} {\bibinfo  {journal} {Procedia Engineering}\
  }\textbf {\bibinfo {volume} {72}},\ \bibinfo {pages} {726} (\bibinfo {year}
  {2014})}\BibitemShut {NoStop}%
\end{thebibliography}%
\end{document}